\documentclass{jfm}
\usepackage{natbib}
\usepackage{amsmath}
\usepackage{amssymb}
\usepackage{graphicx}

\input epsf

\title[Swinging of red blood cells in simple shear]{On the complex dynamics of a red blood cell in simple shear flow}

\author{ Petia M. Vlahovska$^1$, Yuan-nan Young$^2$, Gerrit Danker$^3$, and Chaouqi Misbah$^3$}
\affiliation{
$^1$ Thayer School of Engineering, Dartmouth College, 8000 Cummings
Hall, Hanover NH 03755, USA\\
$^2$ Department Mathematical Sciences, NJIT, Newark, NJ, USA\\
$^3$Laboratoire de Spectrom\'etrie Physique, UMR, 140
avenue de la physique, Universit\'e Joseph Fourier, and CNRS,  38402
Saint Martin d'Heres, France}
\date{24 April 2010}

\newcommand{\el}{{\mu}}
\newcommand{\bnabla}{\nabla}

\newcommand{\deform}{{\mathrm{\dot\gamma}}}
\newcommand{\rot}{{\mathrm{r}}}
\newcommand{\bd}{{\bf d}}
\newcommand{\br}{{\bf r}}
\newcommand{\bv}{{\bf v}}
\newcommand{\bu}{{\bf u}}
\newcommand{\rhat}{{\bf{\hat r}}}
\newcommand{\xhat}{{\bf{\hat x}}}

\newcommand{\bt}{{\bf t}}
\newcommand{\bn}{{\bf n}}
\newcommand{\bx}{{\bf x}}
\newcommand{\visrat}{{\lambda}}
\newcommand{\out}{{\mathrm{ex}}}
\newcommand{\ins}{{\mathrm{in}}}
\newcommand{\mem}{{\mathrm{mm}}}

\newcommand{\Ca}{\mbox{\it Ca}}

\newcommand{\bT}{{\bf T}}
\newcommand{\bI}{{\bf I}}
\newcommand{\bX}{{\bf X}}
\newcommand{\im}{{\mathrm i}}
\newcommand{\refeq}[1]{(\ref{#1})}

\newcommand{\trac}{{\bf{t}}}
\newcommand{\bS}{{\bf y}}
\newcommand{\tracT}{\Theta}
\newcommand{\half} {{\frac{1}{2}}}

\begin{document}

\maketitle
\begin{abstract}
Motivated by the reported peculiar dynamics of a red blood cell in shear flow, we develop  an analytical  theory for the motion of a nearly--spherical fluid particle enclosed by a  visco--elastic incompressible interface in linear flows.   The analysis explains the effect of particle deformability on  the transition from tumbling to swinging  as the shear rate increases.  Near the transition, intermittent behavior is predicted only if the particle has a fixed shape; the intermittency disappears for a deformable particle. Comparison with available phenomenological models based on the fixed shape assumption highlights their physical foundations and limitations.
\end{abstract}

\section{Introduction}
 A membrane--enclosed particle, such as the red blood cell (RBC), exhibits rich dynamics in  flow (see for recent reviews \cite{Abkarian:2008, Vlahovska:CRreview, Guido:CR}).
 Two classic  types of behavior  in steady shear flow are 
(1)~{\it tank--treading} (TT), in which  the RBC  shape is steady and the membrane  rotates as a tank-tread; the cell major axis is tilted with respect to the flow direction and the inclination angle remains fixed in time, and 
 (2) {\it tumbling} (TB), in which the RBC undergoes a periodic flipping motion.
Recently, a new type of motion called  {\it swinging} (SW) has been experimentally observed \cite[]{Abkarian:2007}. In this case, the RBC's  tank--treading is accompanied by small oscillations in the inclination angle. 
Similar phenomenon has been reported for capsules \cite[]{Walter-Rehage-Leonhard:2001, Ramanjuan-Pozrikidis:1998, Navot:1998} and drops covered with adsorbed protein layer \cite[]{Erni:2005}.

The variety of RBC's motions stems from the unique mechanical properties  of the interface. The cell membrane is made of a lipid bilayer attached to an underlying spectrin network.  The lipid bilayer behaves as a two--dimensional incompressible fluid, while the polymer network endows the membrane with shear elasticity. Closed lipid  membranes (vesicles) display TT, TB but no SW \cite[]{Kantsler-Steinberg:2005, Mader:2006, Kantsler-Steinberg:2006, Deschamps:2009, Kantsler-Steinberg:2009}. The inclination angle can oscillate around  the flow direction \cite[]{Kantsler-Steinberg:2006, Mader:2006, Barthes_Biesel-Sgaier:1985}, but this {\it breathing} (VB, also called trembling) motion differs from the swinging observed with RBCs;  in the latter case, swinging does not necessarily involve  shape deformation. 

The type of motion a vesicle or RBC undergoes depends on  the viscosity mismatch between the inner and suspending fluids, applied shear rate, and the non-sphericity of the rest shape.
For a given viscosity ratio and shape,   at low shear rates the RBC tumbles  because elastic tensions immobilize the interface causing the cell to behave as a solid object. At high shear rates the applied stress  overcomes the elastic tensions  and drags the membrane in motion.  As a result, the RBC adopts a steady TT shape whose major axis ``swings''.  This motion originates from the unstressed non-spherical shape of the RBC, in which
 membrane elements at the equator and the poles are not equivalent.  
During one TT--period, an element passes twice through its unstressed position where it releases elastic energy. 
As the shear rate increases, the amplitude  of the angle oscillations decreases.

Phenomenological studies of swinging \cite[]{Abkarian:2007, Skotheim:2007, Kessler-Finken-Seifert:2009, Noguchi:2009b} based on the theory by \cite{Keller-Skalak:1982}, which models the RBC as an ellipsoid of $fixed$ shape,  qualitatively capture the physics of the phenomenon.  
However, the quantitative predictions  match poorly with simulations and experiments.  For example, the model proposed by \cite{Abkarian:2007} agreed with the experimental data only if the shear elastic modulus of the membrane was  assumed to be lower than the generally accepted value.
Furthermore,  \cite{Skotheim:2007} predicted intermittent behavior, for which no evidence was found in the numerical simulations  \cite[]{Kessler-Finken-Seifert:2008, Sui:2008a, Bagchi-Kalluri:2009}. A recent detailed analysis constructed the dynamical phase diagram of the reduced model  \cite[]{Kessler-Finken-Seifert:2009} in both steady and oscillatory shear.  

The purpose of this work is develop a rigorous analytical theory that disposes of the assumptions for fixed ellipsoidal shape and compressible membrane inherent to the studies based on the Keller--Skalak model.
We generalize the theory for the dynamics of a vesicle made of fluid incompressible membranes \cite[]{Misbah:2006, Vlahovska:2007} to include membrane shear elasticity.  Section \ref{formulation}
formulates  the model, Section \ref{vesicle:summary} summarizes the theory for  fluid membranes (vesicles), Section \ref{Effect of shear elasticity} discusses the evolution equations for the particle shape and orientation angle, and Section \ref{results} analyzes the effects of shear elasticity on particle dynamics.

\section{Problem formulation}
\label{formulation}We model the RBC as a closed membrane (``capsule'') with total area $A$.
  The membrane encapsulates a fluid of viscosity
$\visrat\eta$ and it is suspended in a
fluid of viscosity $\eta$; $\visrat$ denotes the viscosity ratio.  Both interior and exterior fluids
are incompressible and Newtonian. The particle has a characteristic
size $R_0$ defined by the radius of a sphere of the same volume.  The
nonsphericity of the rest shape is characterized by an excess
area
\begin{equation}
\label{excess area}
\Delta=A/R_0^2-4 \pi \,.
\end{equation}
The typical value for a RBC is $\Delta\sim 4$.

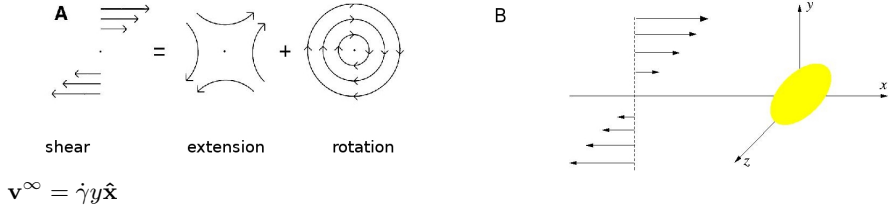
\begin{figure}
\centerline{\includegraphics[height=0.9in]{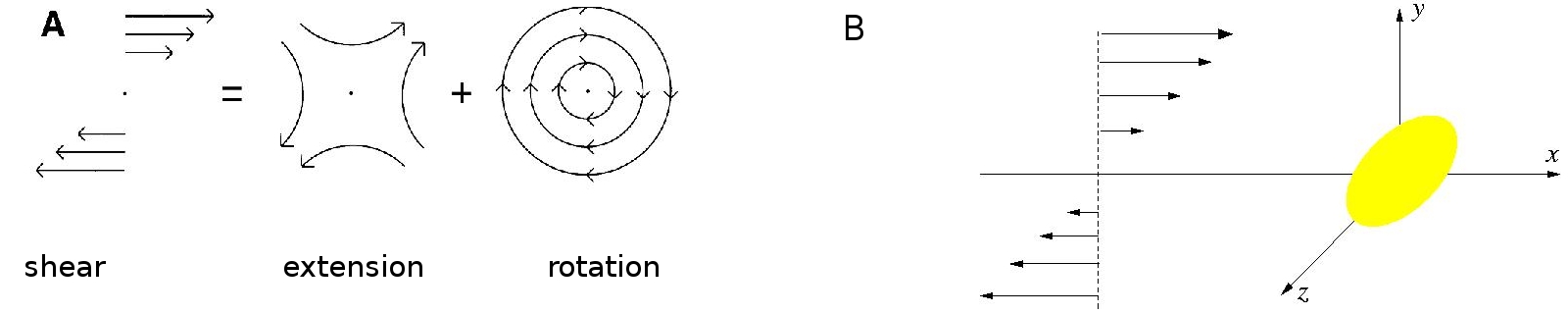}}
\begin{picture}(0,0)(0,0)
\put(20,0){$\bv^\infty=\dot\gamma y \xhat$}
\end{picture}
\caption{\footnotesize A. Sketch of the streamlines of a linear shear flow. The shear flow is a superposition of pure straining flow and rigid body rotation B. A soft particle deforms to an ellipsoid in a simple shear flow that can tank-tread ($\psi=const$) or tumble ($\psi$ continuously increases)}
\label{shear}
\end{figure}

The RBC is placed in a steady two--dimensional linear flow
\begin{equation}
\label{external flow}
\bv^\infty(\br)=\dot\gamma y\cdot \xhat\,,
\end{equation}
where $\dot\gamma$ is the strain rate.
A sketch of the problem is shown in Figure~\ref{shear}.

\subsection{Fluid motion and fluid--membrane coupling}
At the length scale of the micron--size RBC, water is effectively very viscous and creeping--flow conditions prevail. Fluid velocity $\bv^{(\alpha)}$ and pressure $p^{(\alpha)}$ of the interior ($``\alpha=\ins"$) and suspending ($``\alpha=\out"$) fluids obey  the Stokes equations  and the incompressibility condition
\begin{equation}
\label{Stokes equations}
\nabla\cdot\bT^{(\alpha)}=0\,,\quad \nabla\cdot\bv^{(\alpha)}=0\,,
\end{equation}
where  $\bT$  is the bulk hydrodynamic stress
\begin{equation}
\label{stress definition}
\bT^{(\alpha)}= -p^{(\alpha)}\bI+\eta^{(\alpha)}\left[\nabla\bv^{(\alpha)}+(\nabla\bv^{(\alpha)})^\dagger\right]\,.
\end{equation}
$\bI$ denotes the unit tensor and the superscript ${\dagger}$ denotes transpose.
 Far away from the particle, the flow field tends to the unperturbed
external flow $\bv^{\out}\rightarrow \bv^{\infty}$. The velocity field is continuous across the membrane.
However, the hydrodynamic stresses undergo a jump, which is balanced
by membrane surface forces
\begin{equation}
\label{stress balance}
\bn\cdot\left(\bT^\out-\bT^\ins\right)={ \bt}^{\mem}\, ,
\end{equation}
where $\bn$ is the outward unit normal vector and the membrane surface forces ${ \bt}^{\mem} $ are discussed next.

\subsection{Mechanics of flexible fluid membranes}
\label{Membrane mechanics}
The RBC membrane consists of a lipid bilayer attached to protein scaffold.
The lipid bilayer endows the membrane with fluidity, incompressibility and bending rigidity, while
the polymer network gives rise to resistance to shearing.  
The membrane thickness is $5nm$, or about 1/1000 of the cell radius.
Due to the large separation of lengthscales, the membrane can be treated as a two--dimensional surface embedded in a three--dimensional space.

Within the framework of the minimal model \cite[]{Seifert:1997},
the bending resistance gives rise to a surface force density
\begin{equation}
\label{interfacial stress}
{\bt}^\kappa=-\kappa \left(4H^3-4KH+2\nabla_s^2 H\right)\bn\,,
\end{equation}
where $H=(1/2)\nabla\cdot \bn$ and $K=(1/2)[ \left(\nabla\cdot \bn\right)^2+\nabla\bn:\nabla\bn^\dagger]$ are the mean and Gaussian curvatures, and $\kappa$ is the bending modulus.
The surface gradient operator
is defined as $\nabla_{\mathrm{s}}=\bI_\mathrm{s}\cdot\nabla$,
where the matrix $\bI_\mathrm{s}=\bI-\bn\bn$ represents a surface projection.

The bilayer consist of fixed number of lipids, which are optimally packed with fixed area per lipid (under moderate stresses).
As a result, an element of the bilayer membrane
only deforms but can not change its area. Under stress, the membrane develops  tension (a two-dimensional pressure), which
adapts itself to the forces exerted on the membrane in order to keep the local and total area constant \cite[]{Seifert:1999}. This tension is non-uniform along the interface and varies with forcing. The corresponding 
 surface force density is
\begin{equation}
\label{tension stress}
{{\bt}}^\sigma=2\sigma H\bn-\nabla_s \sigma \,.
\end{equation}
where $\sigma$ denotes the local membrane tension.

For lipid bilayers in the fluid phase, the lipids  are free to move within
the monolayer. Therefore,  a pure fluid bilayer membrane is infinitely shearable
\cite[]{Rumy:2006}. However, the polymer network lining the bilayer develops elastic tensions when sheared. Various constitutive laws exist to model the elastic behavior of RBCs and capsule membranes \cite[]{Pozrikidis_capsulebook, Barthes_Biesel:1991}. Assuming a linear elastic behavior, the elastic tractions are given by \cite[]{Barthes_Biesel-Rallison:1981, Edwards-Brenner-Wasan:1991}
\begin{equation}
\label{elstress}
\begin{split}
{\bt}^\mu=&-2(K_A-\mu)(\nabla_s\cdot\bd) H\bn+(K_A-\mu)\nabla_s\nabla_s\cdot\bd +\mu\nabla_s\cdot\left[\nabla_s\bd\cdot \bI_s+\bI_s\cdot \left(\nabla_s\bd\right)^\dagger\right]
\end{split}
\end{equation}
where $\bd$ is the displacement of a material particle of the membrane. $K_A$ is the stretch and $\mu$ is the shear elastic moduli. For RBC $\mu\sim 10^{-6} N/m$ and $K_A\sim200 N/m$ \cite[]{Rumy:2006}.
 We can identify $\alpha_2=K_A-\mu\,,\,\alpha_3=\mu$ in the notation of \cite{Barthes_Biesel-Rallison:1981}. 
 In general,  $\sigma$ in Eq. \refeq{tension stress} includes both tensions arising from shear elasticity and area-incompressibility.
For an area-incompressible membrane the first two terms in Eq. \refeq{elstress} vanish because $\nabla_s\cdot\bd=0$. 




\subsection{Time scales}

Viscous forces exerted by the extensional component of the 
flow act to distort the shape on a time scale
\begin{equation}
\label{deformation time:3}
\tau_\deform=(1+\visrat)\dot\gamma^{-1}\,.
\end{equation}
Several intrinsic relaxation
mechanisms oppose the deformation.
  Bending
stresses work to bring the shape back to its preferred curvature
state; the corresponding time scale is
\begin{equation}
\label{Be-relaxation time}
\tau_{\kappa}=\frac{(1+\visrat)\eta R_0^3}{\kappa}\,.
\end{equation}
Relaxation driven by shear elasticity occurs on a time scale
\begin{equation}
\label{El-relaxation time}
\tau_{\mu}=\frac{(1+\visrat)\eta R_0}{\mu}\,.
\end{equation}
The factor $(1+\visrat)$ reflects the fact that the more viscous fluid controls the dynamics. 
In shear flow, particle rotation away from the extensional axis of the imposed flow effectively decreases  the effect of the straining;  the associated  time scale is
\begin{equation}
\label{rotation}
\tau_\rot=\dot\gamma^{-1}.
\end{equation}
The strength of the relaxation mechanisms that
limit  shape deformation by the flow
is quantified  by the corresponding dimensionless parameters:
the capillary number
\begin{equation}
\label{capillary number:bending}
\Ca_\kappa=\frac{\tau_{\kappa}}{\tau_\deform}\,,
\end{equation}
elastic capillary number
\begin{equation}
\label{capillary number:elastic}
\chi^{-1}=\frac{\tau_{\mu}}{\tau_\deform}=\frac{\eta \dot\gamma a}{\mu},
\end{equation}
and the rotation parameter
\begin{equation}
\label{rotation number}
\frac{t_\rot}{t_\deform}=\visrat^{-1}.
\end{equation}
The interplay of these time scales leads to the complex dynamics of RBCs and vesicles.
The elastic capillary number $\chi^{-1}$ plays the role of a dimensionless shear rate.  For a fluid vesicle $\chi=0$ and the dynamics is relatively insensitive to the shear rate \cite[]{Misbah:2006};
for a given excess area,  the TT-TB transition of a vesicle is controlled by the viscosity ratio $\lambda$.

Henceforth,  all quantities are
non-dimensionalized using $\eta$, $R_0$, and $\dot\gamma$.  Accordingly, the time
scale is $\dot\gamma^{-1}$, the velocity scale is $\dot\gamma R_0$, bulk
stresses are scaled with $\eta\dot\gamma$.

\subsection{Perturbative solution for a nearly--spherical shape}

In order to solve the problem analytically,  we consider a nearly-spherical particle shape, i.e., $\Delta\ll 1$.
In a coordinate system centered at the particle,
the radial position  $r_{\mathrm{s}}$ of the
interface can be represented as
\begin{equation}
\label{perturbation of shape}
r_{\mathrm{s}}=1+f(\theta, \phi, t)\,,
\end{equation}
where $f$ is the deviation of
particle shape from a sphere.
The exact position of the interface
is replaced by the surface of a sphere of equivalent volume,
$r=1$,
and all quantities that are to be evaluated at the interface of the
deformed particle are approximated using a Taylor series expansion. The leading order analysis for fluid vesicles has been done in \cite{Vlahovska:2007}. Here, we modify the solution to include shear elasticity.

In Eq.\ \refeq{perturbation of shape}, the function $f$
representing the shape depends only on angular coordinates.  Thus, it is expanded into series of scalar spherical harmonics
$Y_{jm}$ \refeq{normalized spherical harmonics}
\begin{equation}
\label{expansion of shape in harmonics}
f=\sum_{j=2}^{\infty}\sum_{m=-j}^j f_{jm} Y_{jm}\,,
\end{equation}
In the above equation, the summation starts from nonzero $j=2$ because
 $j=1$ correspond to translation of the center of mass.
The  constraint on fixed total  area  serves to relate the amplitude of the perturbation $f$ and the excess area $\Delta$
\begin{equation}
\label{area constraint}
\begin{split}
\Delta=&\int \frac{r_s^2}{\hat\br\cdot\bn}d \Omega- 4 \pi=\sum_{jm} \frac{\left(j+2\right)\left(j-1\right)}{2} f_{jm}f^*_{jm}+O(f^3)
\end{split}
\end{equation}
where $\hat\br$ denotes the unit radial vector and the sum over $j$ starts from 2, $|m|\le j$ and $f^*_{jm}=(-1)^mf_{j-m}$. Since the rest shape of the particle is characterized by small excess area $f\sim \Delta^{1/2}\ll 1$. 

\section{Dynamics of a nearly--spherical closed fluid membrane  in a shear flow}
\label{vesicle:summary}
The theory for vesicles, $\chi=0$, is well developed \cite[]{Misbah:2006, Vlahovska:2007, Lebedev:2008, Danker:2007b, Kaoui-Farutin-Misbah, Schwalbe}
Here we summarize the main results.
In shear flow, a vesicle deforms into an ellipsoid and, hence, its  shape is specified by the $j=2$ spherical harmonics. In the flow plane $x-y$, the vesicle shape $f$ is characterized by two components, $f_{2\pm2}$ corresponding to deformation along the flow  axis $x$ and the straining axis $x=y$. The out-of-plane deformation along the vorticity  $z$ axis is described by the $f_{20}$ mode. For simplicity, $f_{2\pm1}$ modes are neglected.
Instead of shape modes, the vesicle dynamics can be more conveniently described in terms of the orientation angle, $\psi$, and  $R$, which  measures the ellipticity of the vesicle contour in the $x-y$ plane \cite[]{Misbah:2006}
\begin{equation}
f_{2\pm2}=R \exp(\mp2\im \psi)\,.
\end{equation}
The $f_{20}$ is slaved to the $f_{2\pm2}$ modes and it is determined from the area constraint \refeq{area constraint}
\begin{equation}
\label{f20}
f_{20}=\left[\frac{\Delta}{2}-2 f_{22} f_{2-2}\right]^{1/2}=\left[\frac{\Delta}{2}-2 R^2\right]^{1/2}\,.
\end{equation}
The evolution equations for the shape and orientation of a fluid membrane vesicle in a simple shear flow are \cite[]{Misbah:2006}
\begin{equation}
\label{psidot}
\frac{\partial \psi}{\partial t}=-\frac{1}{2}+\frac{h}{2 R(t)}\cos\left[2\psi(t)\right]\,,
\end{equation}
\begin{equation}
\label{Rdot}
\frac{\partial R}{\partial t}= h\left(1-4 \frac{R(t)^2}{\Delta}\right)\sin\left[2\psi(t)\right]\,,
\end{equation}
where $h=4\sqrt{30\pi}/(23 \lambda+32)$.  
Unlike drops and initially spherical capsules,  the vesicle motion described by equations \refeq{psidot} and \refeq{Rdot} is nonlinear (even at leading order) and independent of the elastic properties of the interface. This  is due to fact that the interfacial stresses are dominated by the tension that arises from the surface--incompressibility.

This set of coupled nonlinear equations  has a stable fixed point corresponding to  the tank-treading state ($R^*=\sqrt{\Delta}/{2}\,, \psi^*=\arccos{\sqrt{\Delta}/{h}}$) and a closed orbit centered at ($\psi^*=0\,, R^*=h$) describing the breathing mode.  Tumbling corresponds to no equilibrium point. 
The TT fixed point loses stability at
a critical viscosity ratio
\begin{equation}
\label{crit viscosity}
\visrat_c=-\frac{32}{23}+\frac{120}{23}\sqrt{\frac{2\pi}{15 \Delta}}\,.
\end{equation}

If there is no deformation along the vorticity direction, i.e.,  $f_{20}=0$ at all times, Eq. \refeq{f20} implies that  $R$ remains constant and equal to its maximum value $\sqrt{\Delta}/2$.
This situation resembles  the Keller-Skalak model: the vesicle shape is a fixed ellipsoid and the vesicle dynamics is described only by the variations of the angle $\psi$ (note, however, that unlike the Keller-Skalak solution, our velocity field is strictly area--incompressible).  In this case the bifurcation is from TT to TB.  

If $f_{20}\ne 0$,  the transition is from TT to VB.
In the breathing (VB) mode, the vesicle undergoes periodic shape deformations along the vorticity direction. As a result, the vesicle appears to tremble in the flow direction. 

As already discussed by  \cite{Lebedev:2008} and  \cite{Kaoui-Farutin-Misbah},  within the leading order theory \refeq{psidot} and \refeq{Rdot} the TB and VB modes coexist, and the mode selection is determined by the initial conditions.  


\section{Effect of the shear elasticity}
\label{Effect of shear elasticity}

The inclusion of solid--elastic deformation in the fluido--dynamic problem is, in general, not a trivial task \cite[]{Barthes_Biesel:1980}. Since 
 fluids have no memory,  their motion is described in a fixed laboratory frame (Eulerian approach).
However,  membrane elastic stresses depend on the displacement of material particles, i.e., the membrane deformation has to be, in principle,  described in a material (Lagrangian) frame.  For example, at steady tank-treading state, the deformation at a fixed Eulerian point  is constant in time unlike that of a fixed material point  because material elements rotate.
Coupling the membrane deformation and fluid motion
 requires transformation between the Lagrangian and the Eulerian representations. A nice discussion of this problem is given  by \cite{Barthes_Biesel-Rallison:1981}; we closely follow their approach.

\subsection{Reference unstressed shape}
The position of a membrane element in the unstressed configuration is labeled by $\bX$. At time $t$,  its position in the Eulerian frame is $\bx(\bX,t)$. The element displacement is 
\begin{equation}
\bd=\bx(r,\theta,\phi,t)-\bX(r,\theta,\phi,t)
\end{equation}
The motion of the interface is parametrized with respect to a sphere
\begin{equation}
\bd(t)=\bd_t+r_s(t) \rhat \,.
\end{equation}
where $\bd_t$ is tangential to a sphere. Since the interface is area--incompressible,   the radial and tangential displacement components  are not independent. Accordingly, the radial displacement is sufficient to describe the motion of a material point; the tangential displacement is determined from the condition $\nabla_s\cdot \bd=0$.
The unstressed shape is assumed to coincide with the initial shape and it is specified by 
\begin{equation}
r_{ref}(t)=1+g_{jm}Y_{jm}(\theta,\phi)\,.
\end{equation}
The imposed shear flow consists of straining flow and rigid body rotation, which does not generate deformation. Accordingly,
the distortion of a material element needs to be defined relative to a rotating unstressed configuration. 
 At leading order,
 the reference unstressed configuration rotates with the flow vorticity
\begin{equation}
\frac{\partial g_{2m}}{\partial t}=\im \frac{m}{2}g_{2m}\,.
\end{equation}

\subsection{Evolution equations}
The solution follows standard steps \cite[]{Vlahovska:2007} and it is summarized in Appendix \ref{app:solution}.
We obtain that the shear elasticity modifies the evolution equations \refeq{psidot} and \refeq{Rdot} as follows
\begin{equation}
\label{psidot1}
\begin{split}
\frac{\partial \psi}{\partial t}=&-\frac{1}{2}+\frac{h}{2 R(t)}\cos\left[2\psi(t)\right]+\chi\frac{2hr_0}{R(t)\sqrt{30\pi}}\sin\left[2\phi(t)-2\psi(t)\right]
\end{split}
\end{equation}
\begin{equation}
\label{Rdot1}
\begin{split}
\frac{\partial R}{\partial t}=& h\left(1-4 \frac{R(t)^2}{\Delta}\right)\sin\left[2\psi(t)\right]+\chi\frac{4h}{\Delta\sqrt{30\pi}}\left\{r_0\left(\Delta-4 R(t)^2\right)\cos\left[2\phi(t) -2\psi(t)\right]\right.\\
&\left.-2 R(t) f_{20}(t) g^0_{20}\right\}
\end{split}
\end{equation}
where 
\begin{equation}
\label{f20b}
f_{20}(t)=\sqrt{\Delta/2-2R(t)^2}
\end{equation}and $r_0\ne 0$ defines a non-spherical reference shape
\begin{equation}
g^0_{2\pm2}=r_0 \exp(\mp2\im \psi_0) \qquad g^0_{20}=\sqrt{\Delta/2-2r_0^2}\,,
\end{equation}
where $ g^0_{jm}=g_{jm}(t=0)$.
The tank--treading frequency is
\begin{equation}
\label{ttangle}
\frac{\partial \phi}{\partial t}=-\frac{1}{2}\,,
\end{equation}
where $\phi$ is the angle between the position vector of a material particle and the flow direction.
If the reference (unstressed) configuration of the vesicle is a sphere, both $r_0=0$ and $g^0_{20}=0$. In this case the evolution equations become independent of the shear elasticity,
similarly to the way bending elasticity scales out  from the evolution equations for the vesicle dynamics \cite[]{Olla:2000}.

\section{Results and discussion}
\label{results}
In this section we first discuss the reduced model corresponding to a non-deformable ellipsoid  in order to compare our work to earlier studies based on the Keller-Skalak's theory.
 After that we proceed to analyze the more general case of a deformable particle.

\subsection{Dynamics of a particle with a fixed ellipsoidal shape}
The evolution equations \refeq{psidot1} and \refeq{Rdot1} reduce to a shape-preserving model, if there is no out-of-the-shear-plane deformations, $f_{20}=const$. Because the evolution of the $f_{20}$ mode is slaved to the $2\pm2$ modes (as seen from \refeq{f20b}), this condition  can be strictly enforced only if $f_{20}=0$ and $g^0_{20}=0$. This implies that the ellipticity of the vesicle contour  is constant
\begin{equation}
R(t)=r_0=\sqrt{\Delta}/2\,.
\end{equation}
The particle dynamics is described by only one variable, the inclination angle $\psi$
\begin{equation}
\label{psidot-el}
\frac{\partial \psi}{\partial t}=-\frac{1}{2}+\frac{h}{\sqrt{\Delta}}\cos\left(2\psi\right)-\chi \frac{2h}{\sqrt{30\pi}} \sin\left(t+2\psi-2\psi_0\right)\,.
\end{equation}

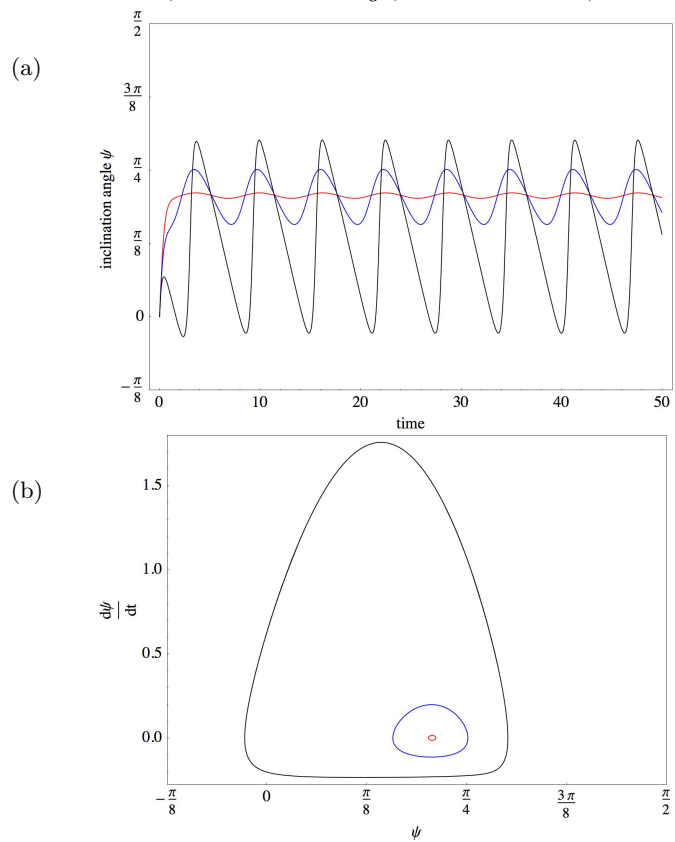
\begin{figure}
\centerline{\includegraphics[width=3in]{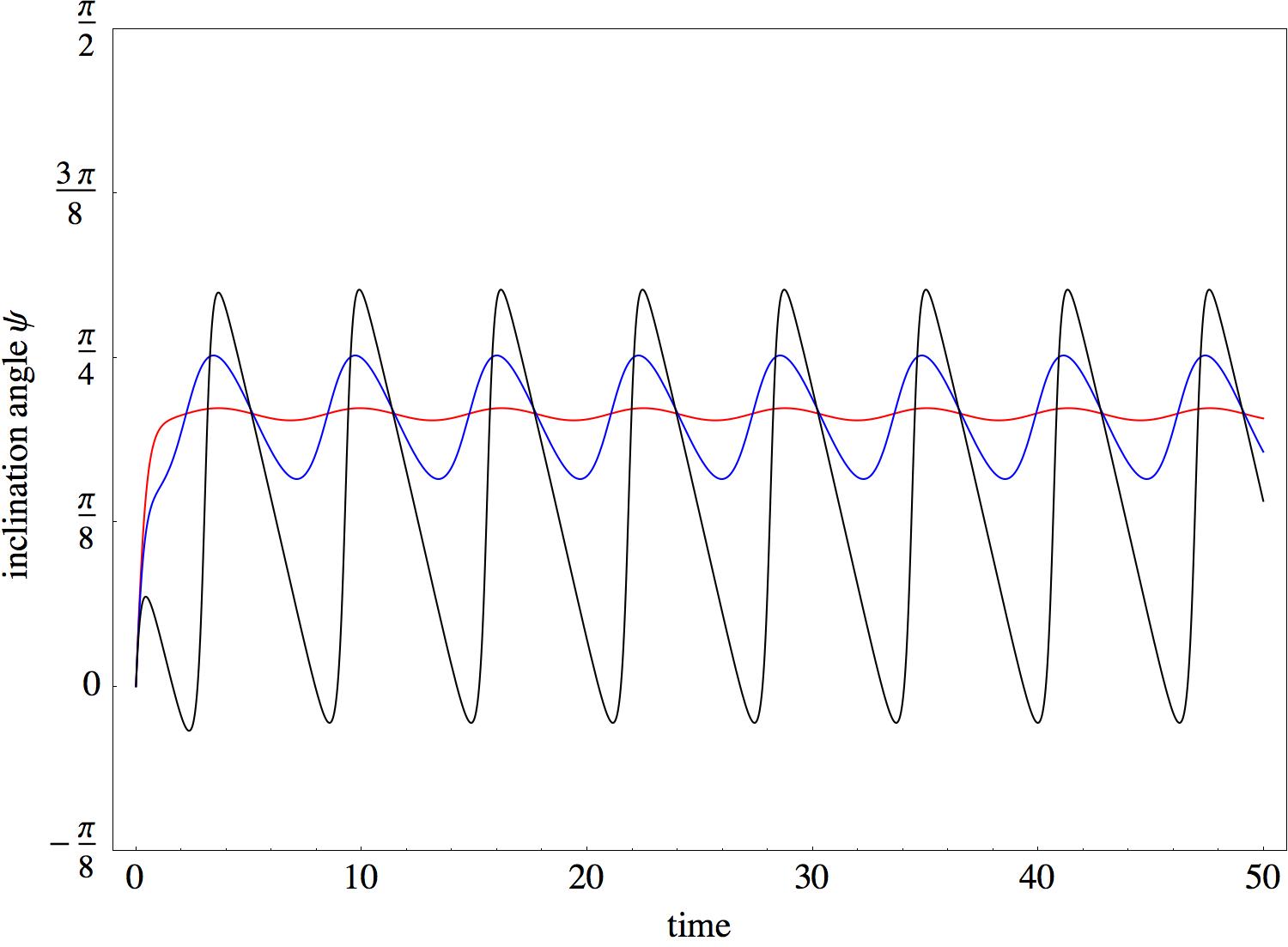}}
\centerline{\includegraphics[width=3in]{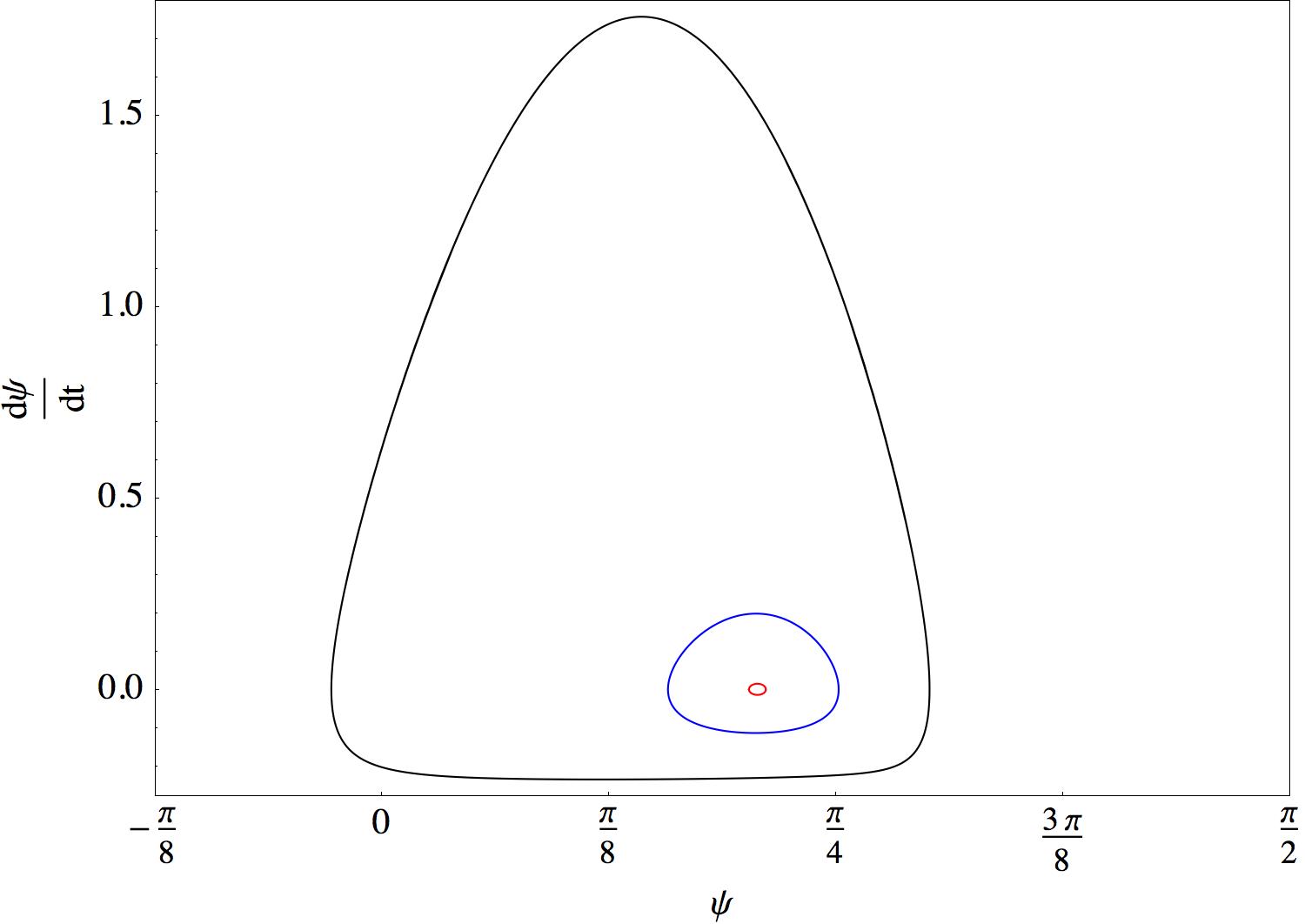}}
\begin{picture}(0,0)(0,0)
\put(50,300){(a)}
\put(50,140){(b)}
\end{picture}
      \caption{\footnotesize  Swinging motion: (a) the inclination angle oscillates in time. (b) the limit cycle in the phase portrait. The amplitude of the oscillations increases with the elasticity number $\chi=1$ (red) , $\chi=10$ (blue) $\chi=29$ (black).  The other parameters are: $\Delta=0.02$ and $\visrat=5$ .}
      \label{fig1}
\end{figure}

Examples of the angle evolution are shown in Figures \ref{fig1}-\ref{fig5}.  At high shear rates (small $\chi$), the particle is swinging; the inclination angle oscillates around a positive value as shown in Figure \ref{fig1}.a. Eq.\refeq{psidot-el} has a stable limit cycle as seen in Figure \ref{fig1}.b. As the shear rate decreases ($\chi$ increases) the amplitude of the oscillations increase and the radius of the limit cycle also increases. Eventually  the limit cycle loses stability and the particle begins to tumble. The tumbling dynamics is illustrated in Figures \ref{fig2}.a and \ref{fig2}.b.  Increasing the viscosity ratio, at a fixed elasticity, also causes the limit cycle to disappear in favor of tumbling because increasing the viscosity ratio decreases the mean inclination angle. However 
the swinging amplitude is relatively insensitive to the viscosity ratio, as seen in Figures \ref{fig3}.a and \ref{fig3}.b.

At intermediate values of the elasticity parameter and high viscosity ratios, the particle exhibits intermittent behavior, which was first reported by  \cite{Skotheim:2007}.  Figures \ref{fig4}.a-\ref{fig5}.c illustrate the complexity of the intermittent dynamics.
\begin{figure}
\centerline{\includegraphics[width=3in]{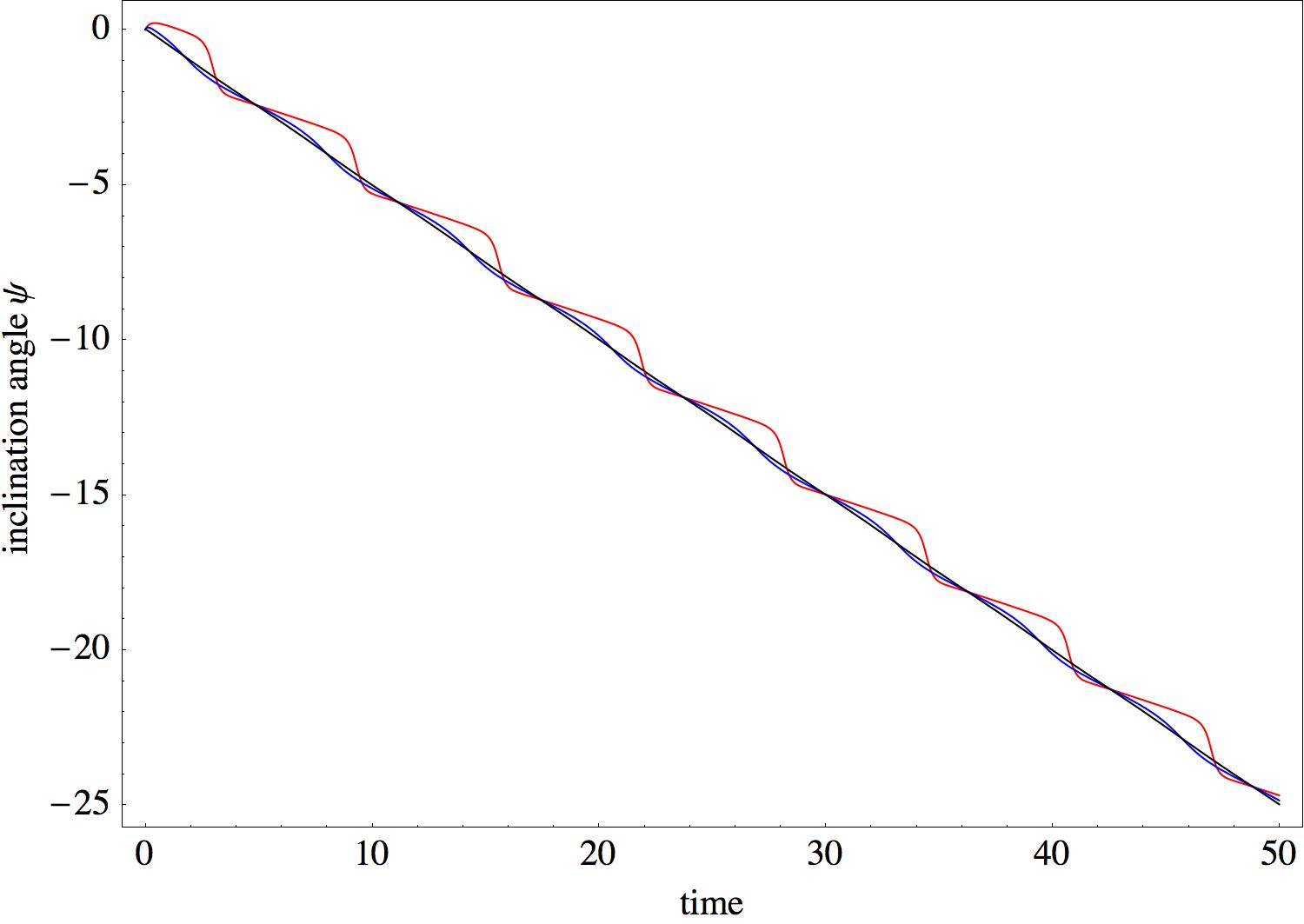}}
\centerline{\includegraphics[width=3in]{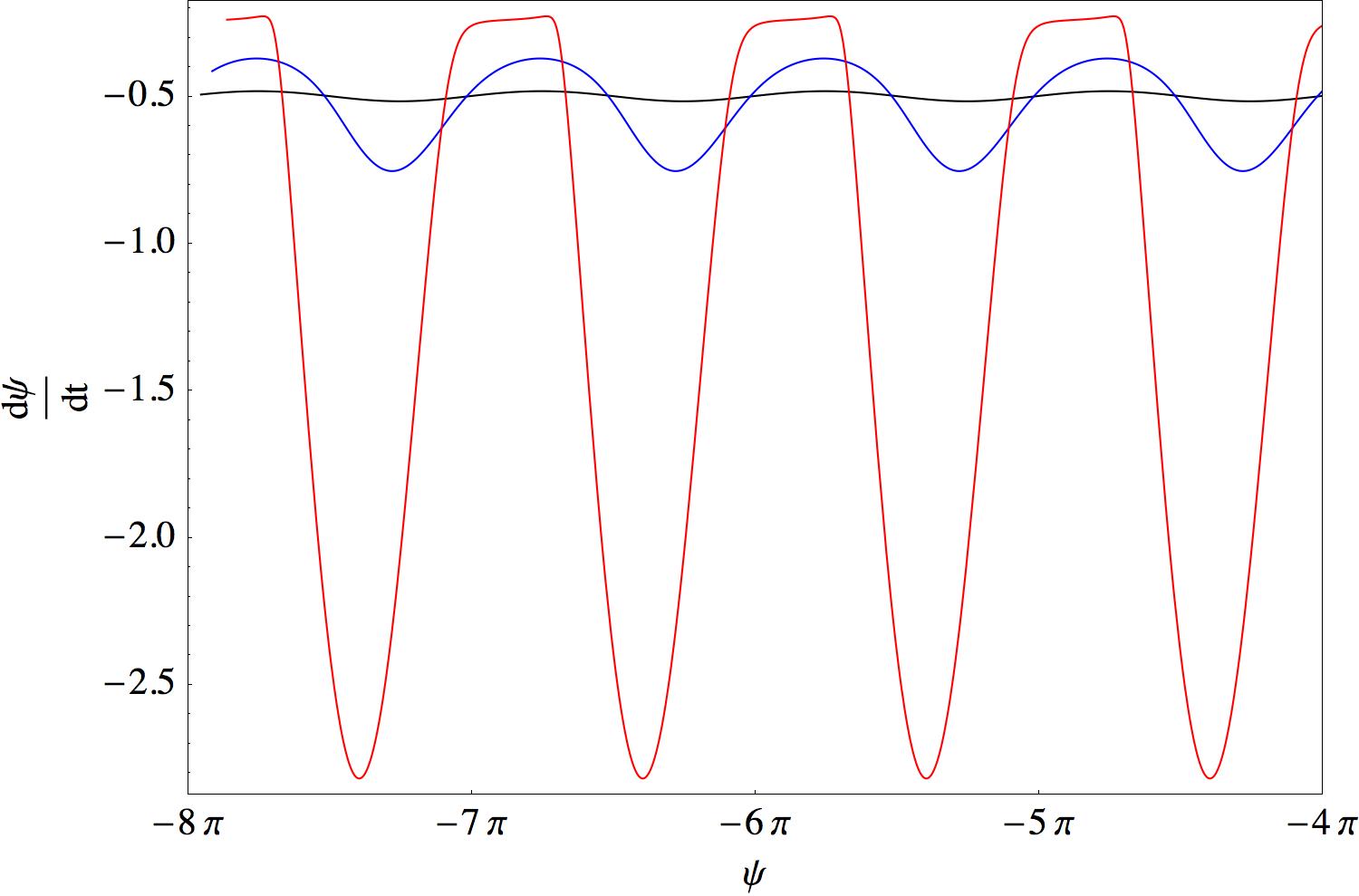}}
\begin{picture}(0,0)(0,0)
\put(50,300){(a)}
\put(50,140){(b)}
\end{picture}
      \caption{\footnotesize  Tumbling motion: (a) the inclination angle decreases continuously with time. (b) the phase portrait shows the angular velocity as a function of the inclination angle. Close to the transition the angular velocity is nonuniform. As the elasticity increases the limiting behavior of rigid ellipsoid is approached. Parameters: $\chi=30$ (red) , $\chi=100$ (blue), and  $\chi=1000$ (black). The excess area is $\Delta=0.02$ and viscosity ratio is $\visrat=5$. }
      \label{fig2}
\end{figure}
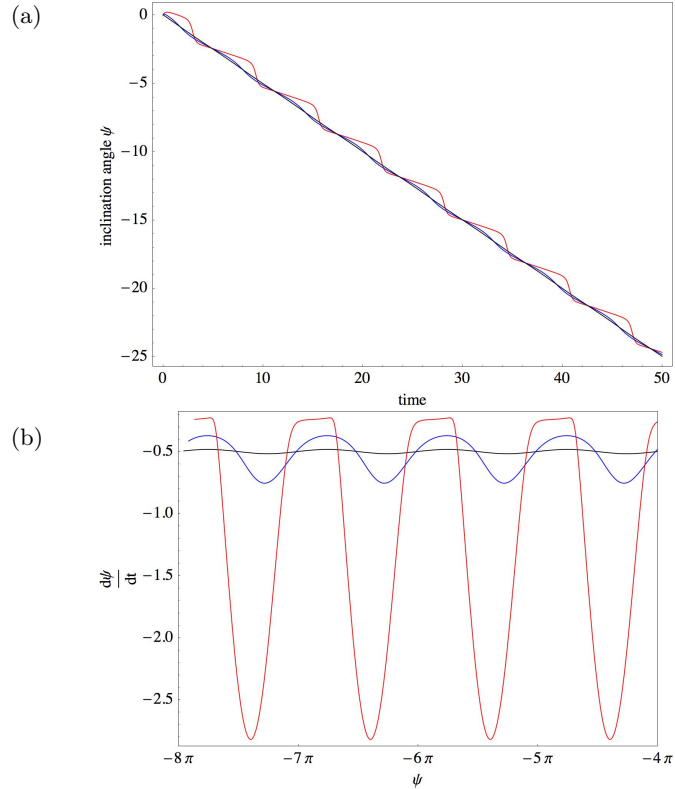
\begin{figure}
\centerline{\includegraphics[width=3in]{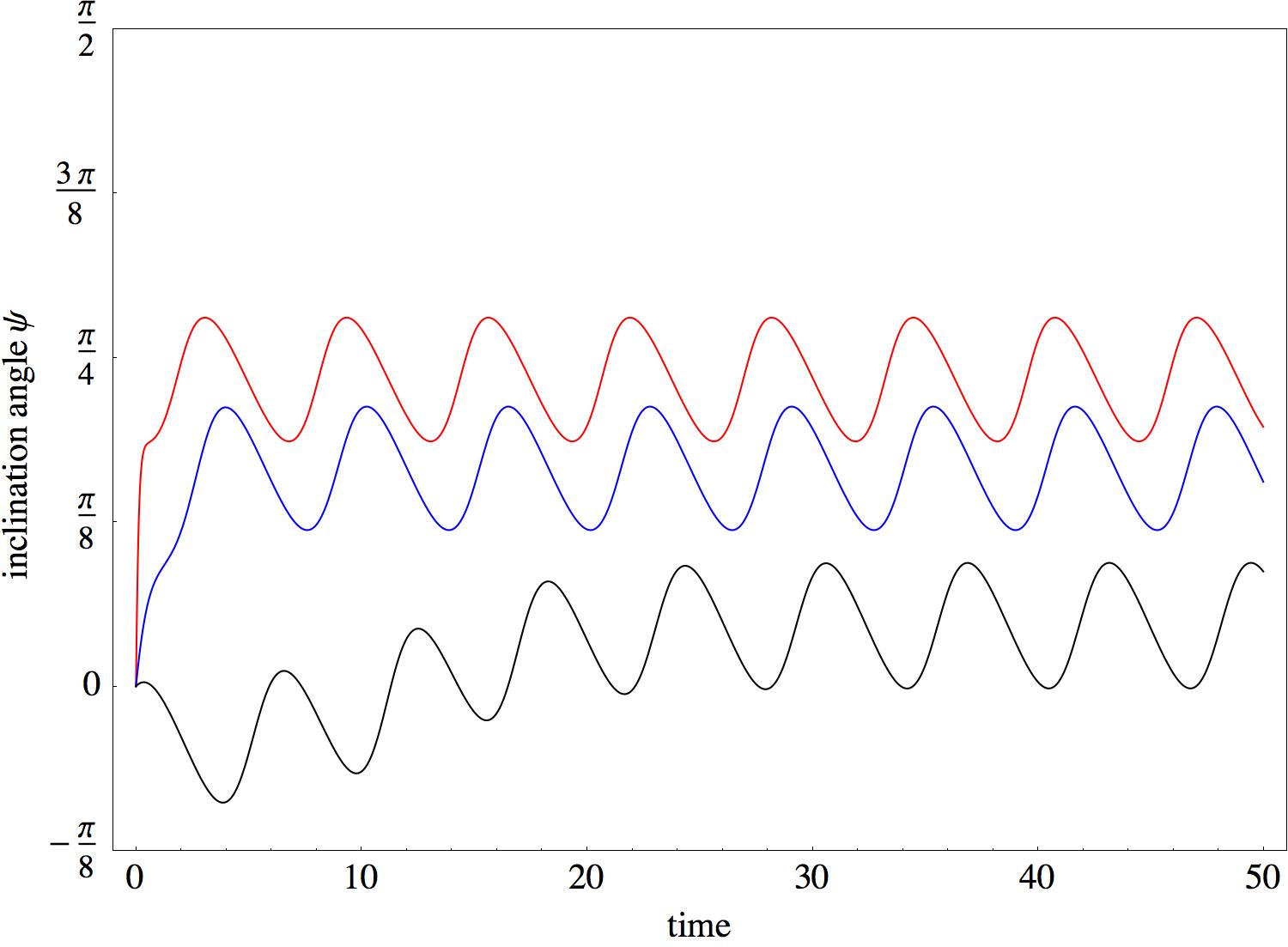}}
\centerline{\includegraphics[width=3in]{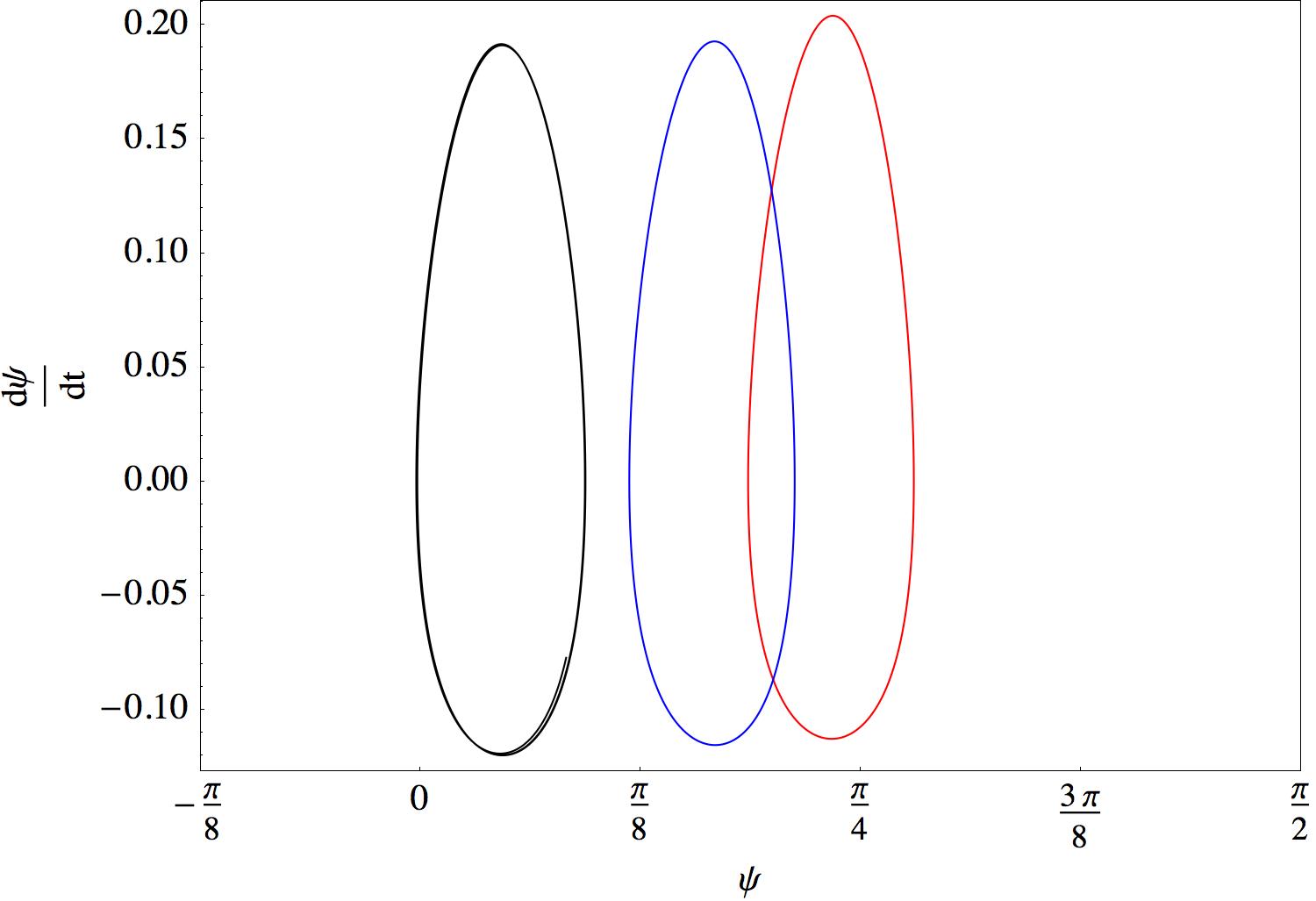}}
\begin{picture}(0,0)(0,0)
\put(50,300){(a)}
\put(50,140){(b)}
\end{picture}
      \caption{\footnotesize Effect of viscosity ratio on the swinging motion. (a) time evolution of the inclination angle. (b) phase portrait. Parameters: $\visrat=1$ (red) , $\visrat=10$ (blue), and $\visrat=20$ (black).The excess area is $\Delta=0.02$ and elastic capillary number is $\chi=10$.}
      \label{fig3}
\end{figure}
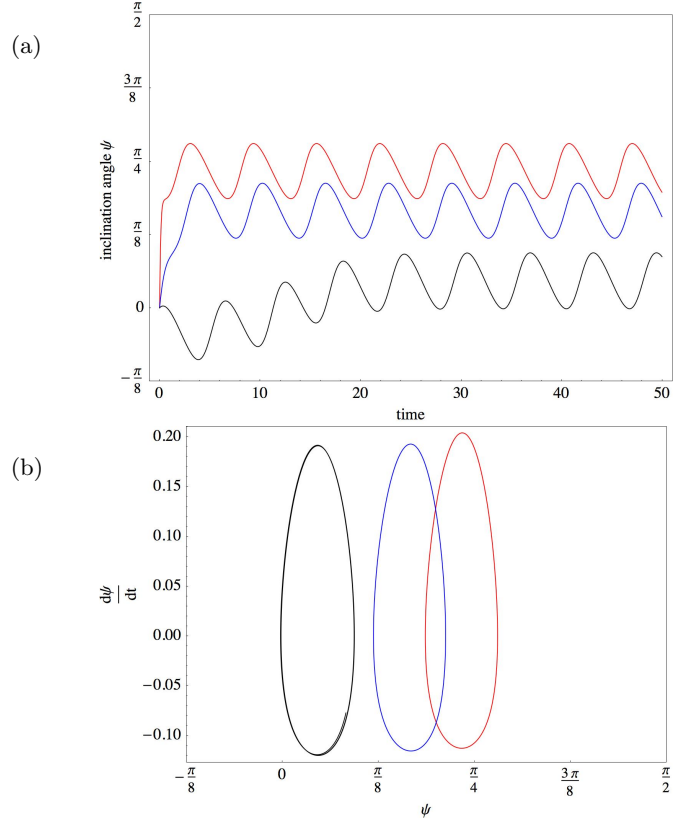
\begin{figure}
\centerline{\includegraphics[width=3in]{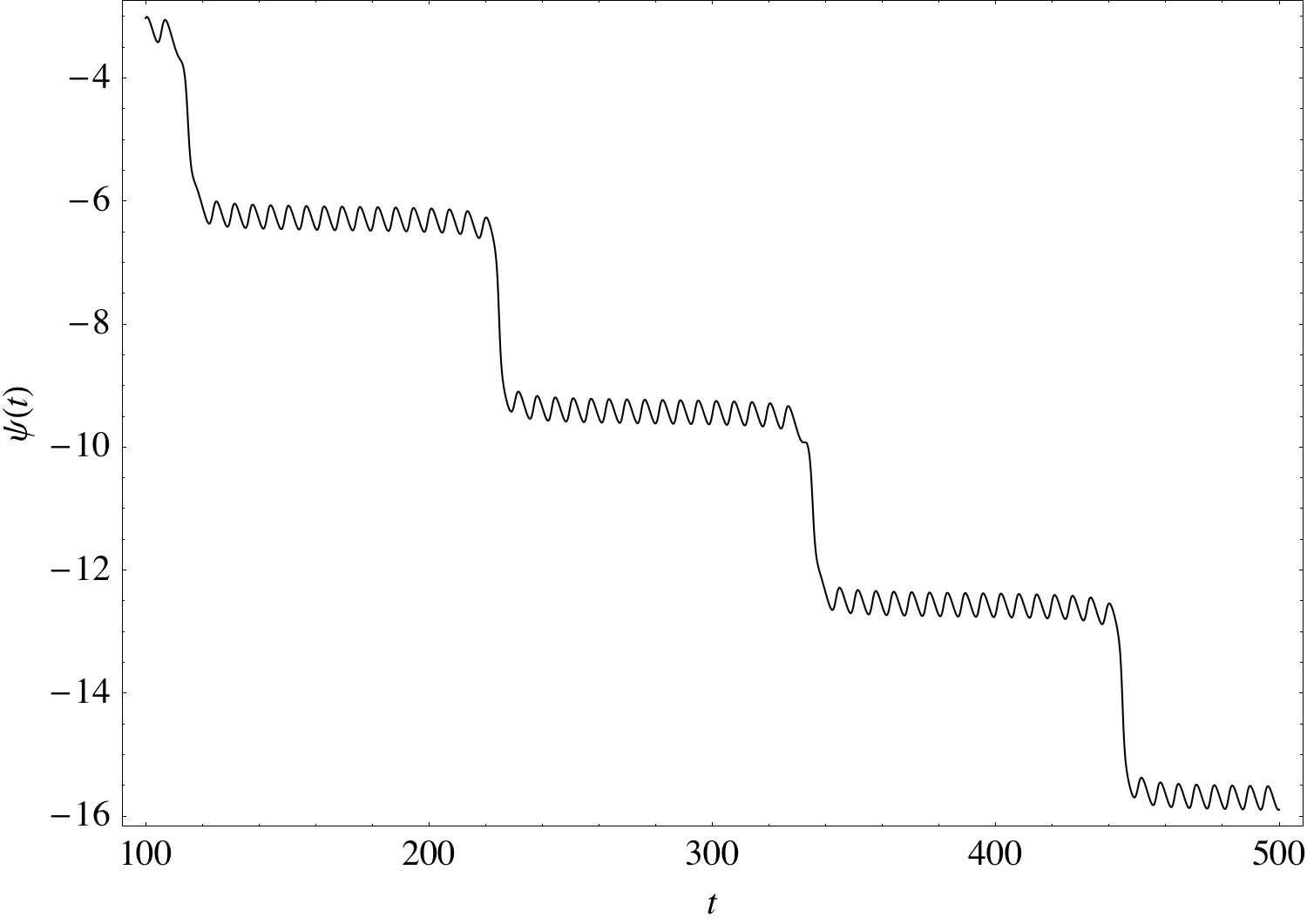}}
\centerline{\includegraphics[width=3in]{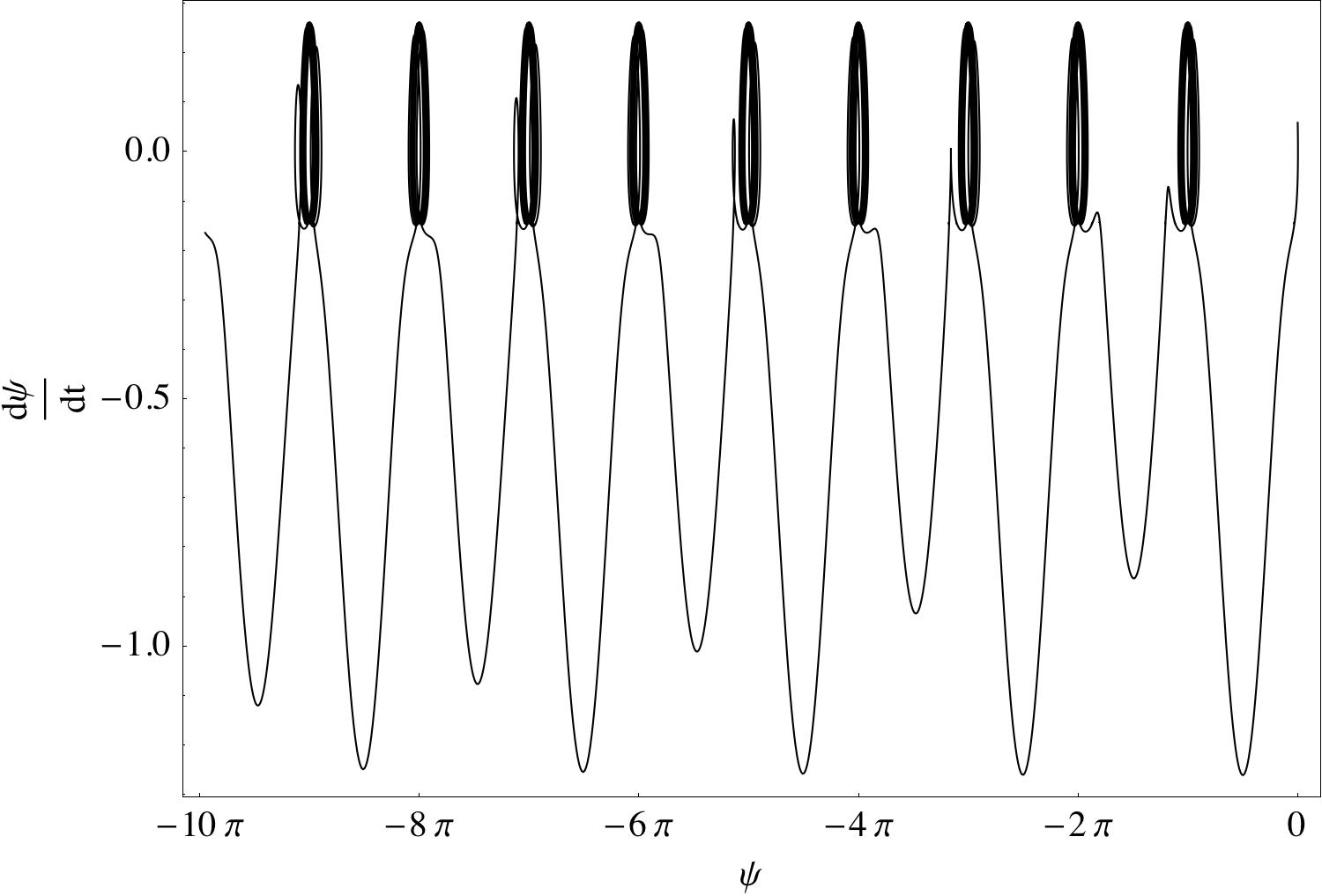}}
\centerline{\includegraphics[width=3in]{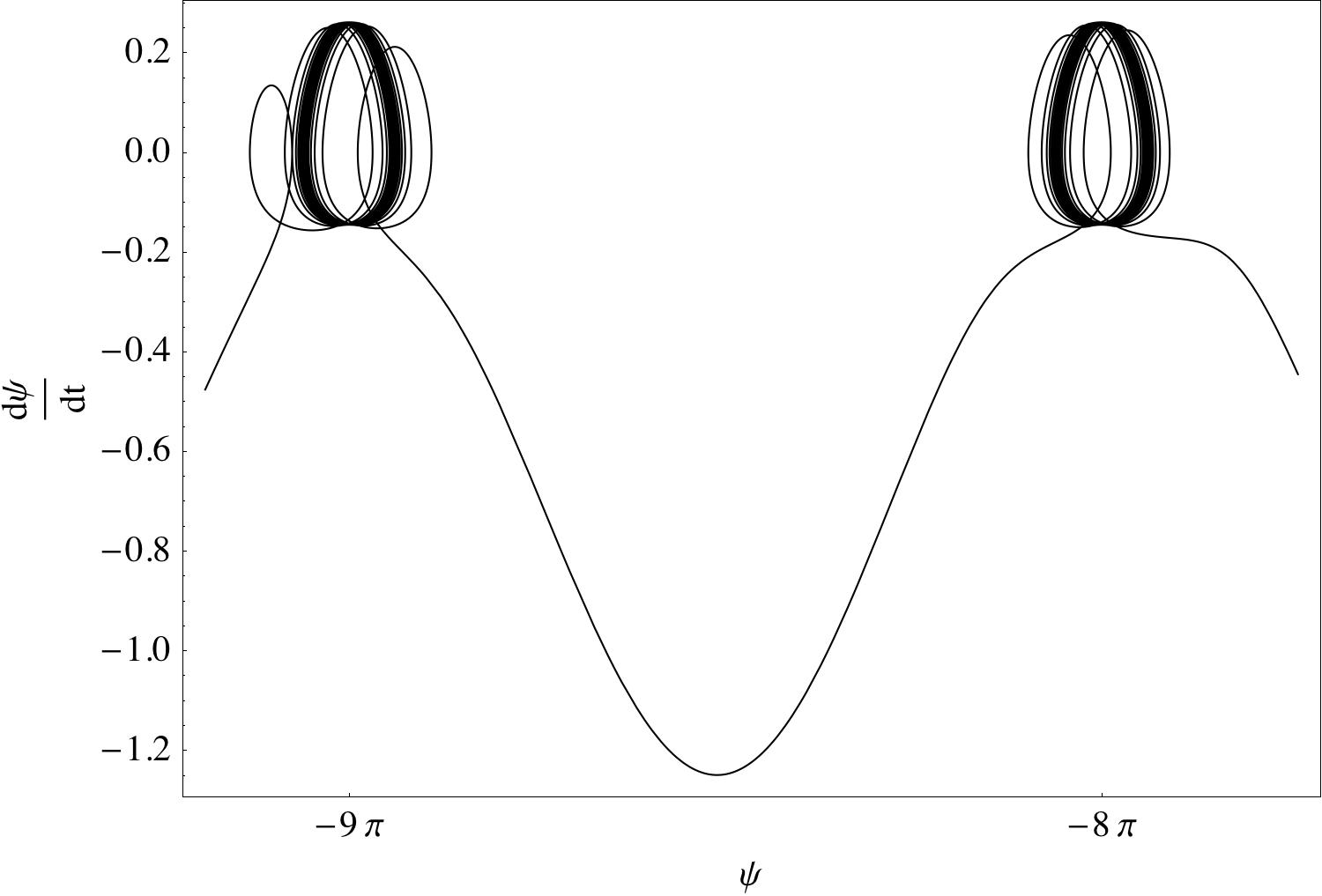}}
\begin{picture}(0,0)(0,0)
\put(50,450){(a)}
\put(50,280){(b)}
\put(50,150){(c)}
\end{picture}
      \caption{\footnotesize  Intermittent dynamics: (a) time evolution of the inclination angle. (b) phase portrait. (c) zoom into the phase portrait. Parameters: $\Delta=0.02$, $\chi=12.5$, $\visrat=20$ .}
      \label{fig4}
\end{figure}
\begin{figure}
\centerline{\includegraphics[width=3in]{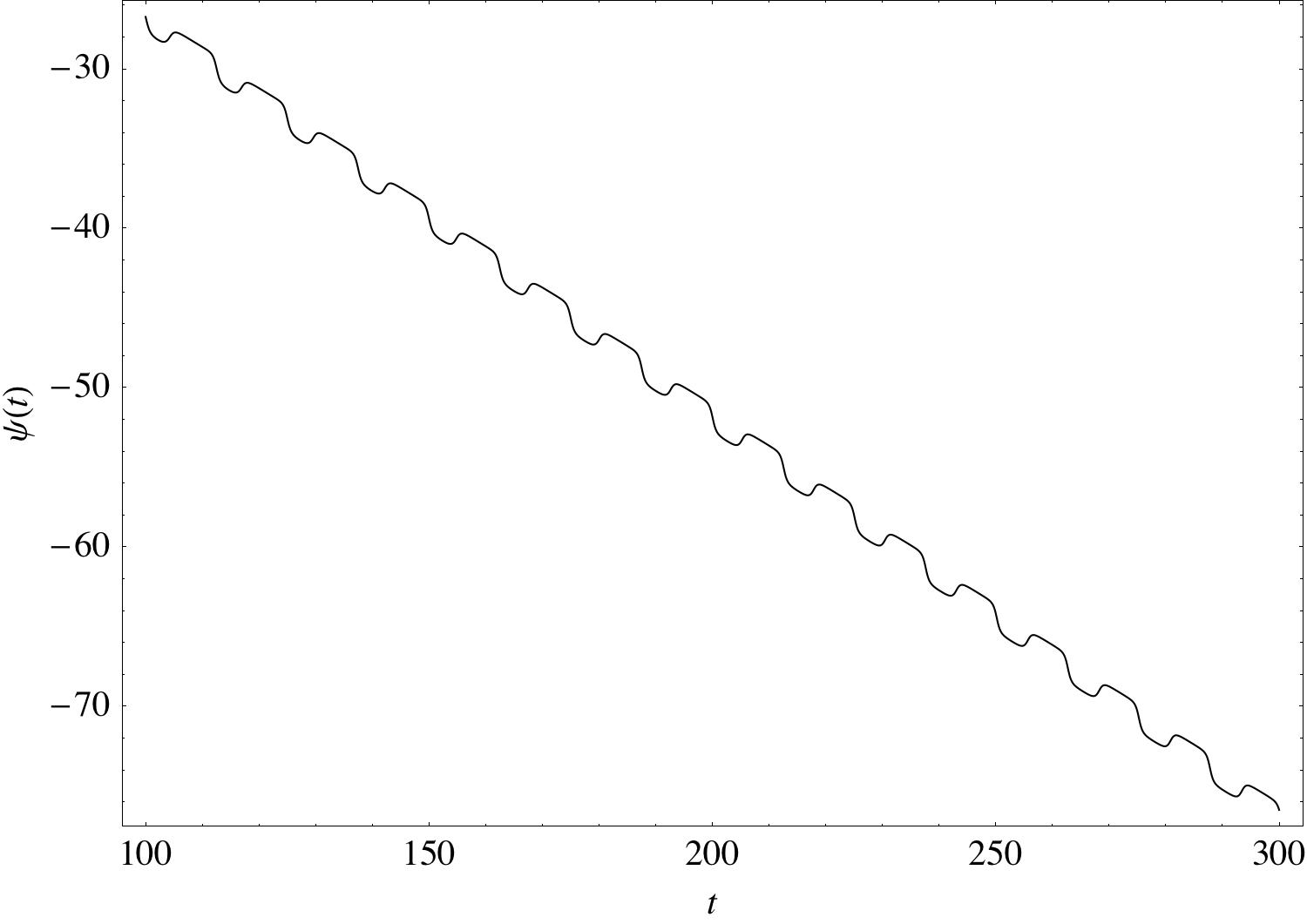}}
\centerline{\includegraphics[width=3in]{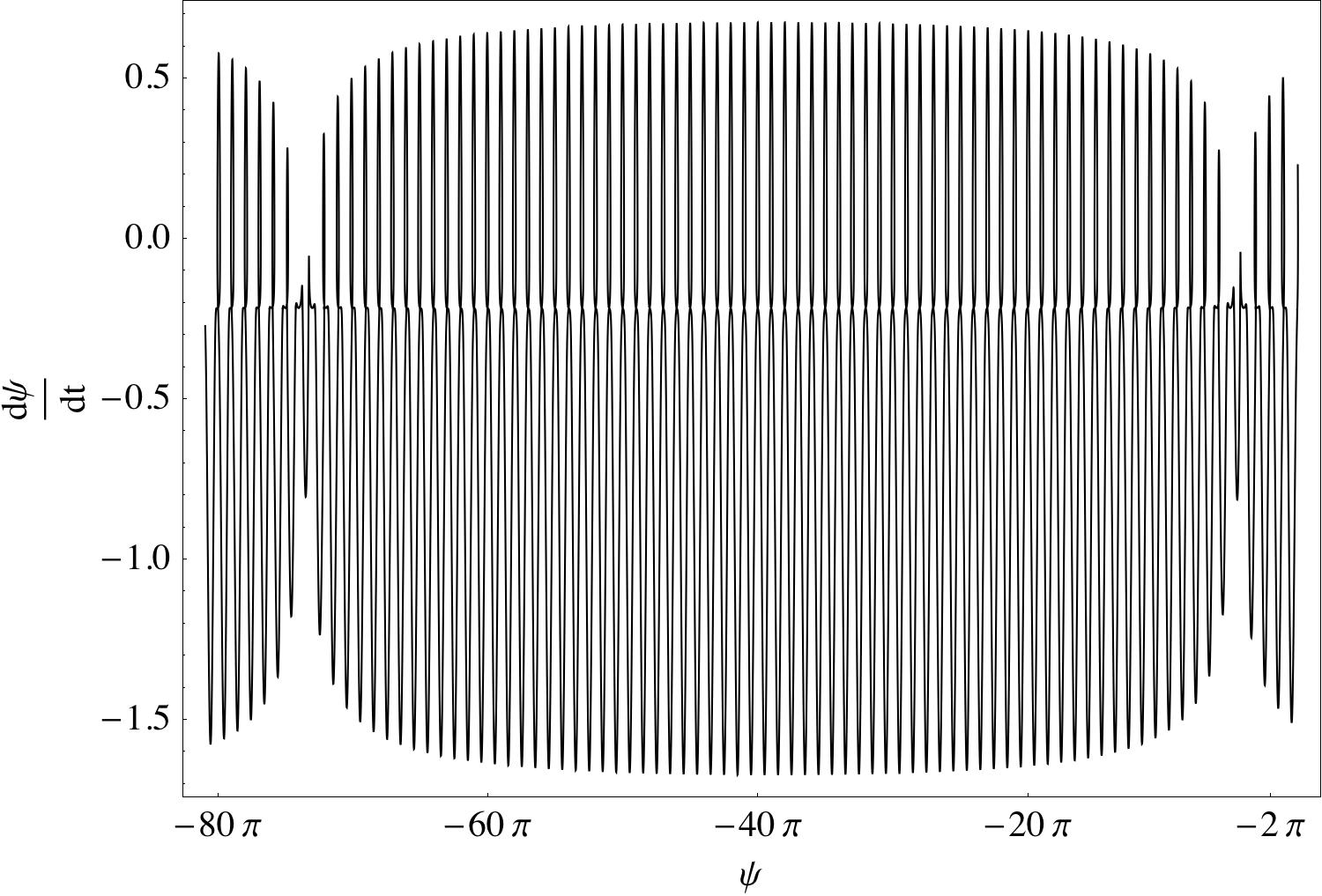}}
\centerline{\includegraphics[width=3in]{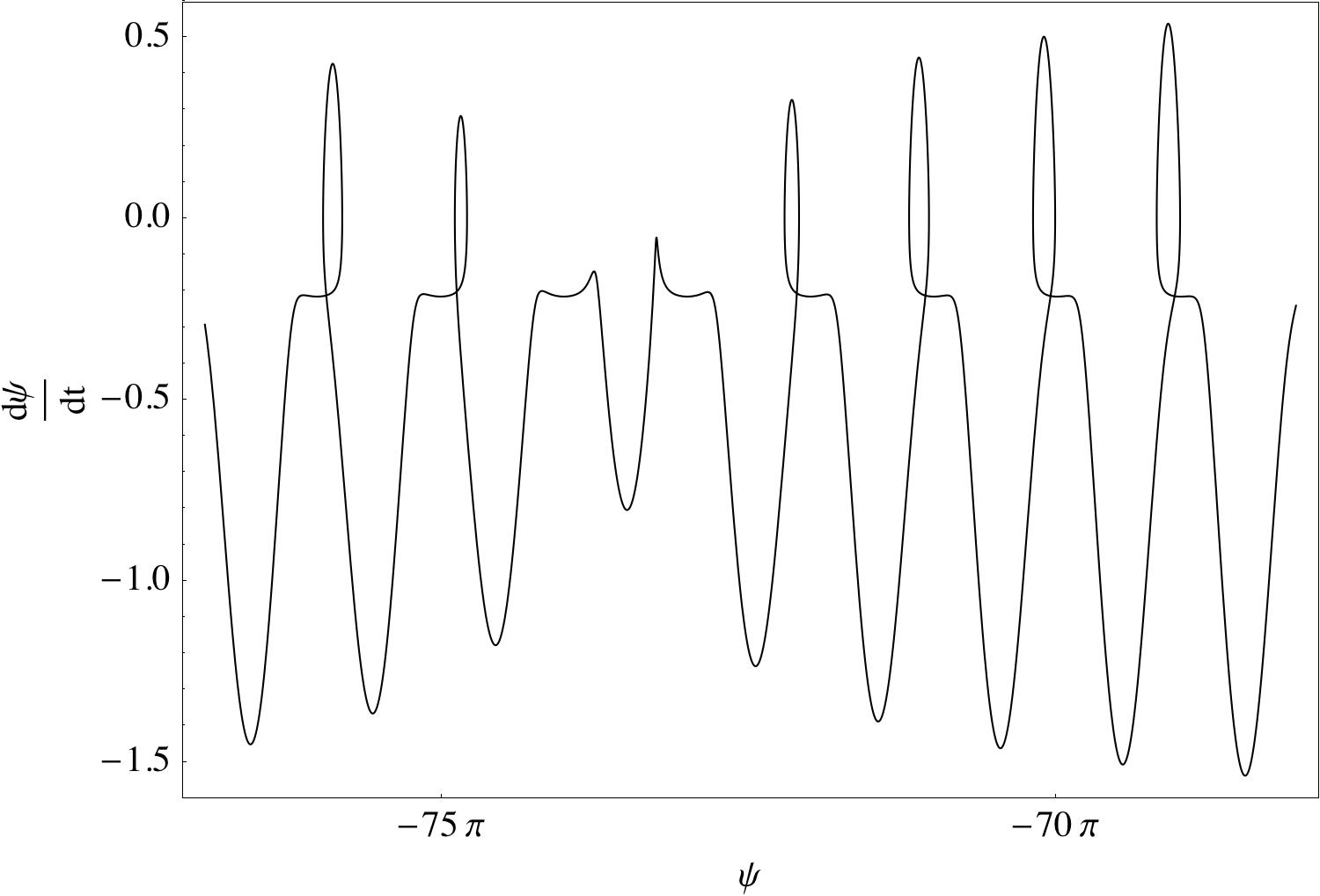}}
      \caption{\footnotesize  Intermittent dynamics: (a) time evolution of the inclination angle. (b) phase portrait. (c) zoom into the phase portrait. Parameters: $\Delta=0.02$, $\chi=21$, $\visrat=15$ .}
      \begin{picture}(0,0)(0,0)
\put(50,450){(a)}
\put(50,280){(b)}
\put(50,150){(c)}
\end{picture}
      \label{fig5}
\end{figure}

In order to compare with earlier works, we have considered an ellipsoid with excess area $\Delta=0.02$, which corresponds to a prolate ellipsoid with axes 0.9, 0.9 and 1. For this small excess area, the Keller-Skalak model ($\chi=0$) is in excellent agreement with the exact theory \refeq{psidot-el} 
and only slightly overestimates the viscosity ratio at the transition from TT to TB.  However, when the elasticity is nonzero there is discrepancy between our results and those of \cite{Skotheim:2007}, mainly because it is impossible to relate their phenomenological parameter $U_e$ to the elastic capillary number $\chi$. Our equations, however, agree with those derived by  \cite{Kessler-Finken-Seifert:2009} (see Appendix \ref{kessler}). 

\subsubsection{Phase boundary}
\label{phaseb}
\begin{figure}
\centerline{\includegraphics[width=3in]{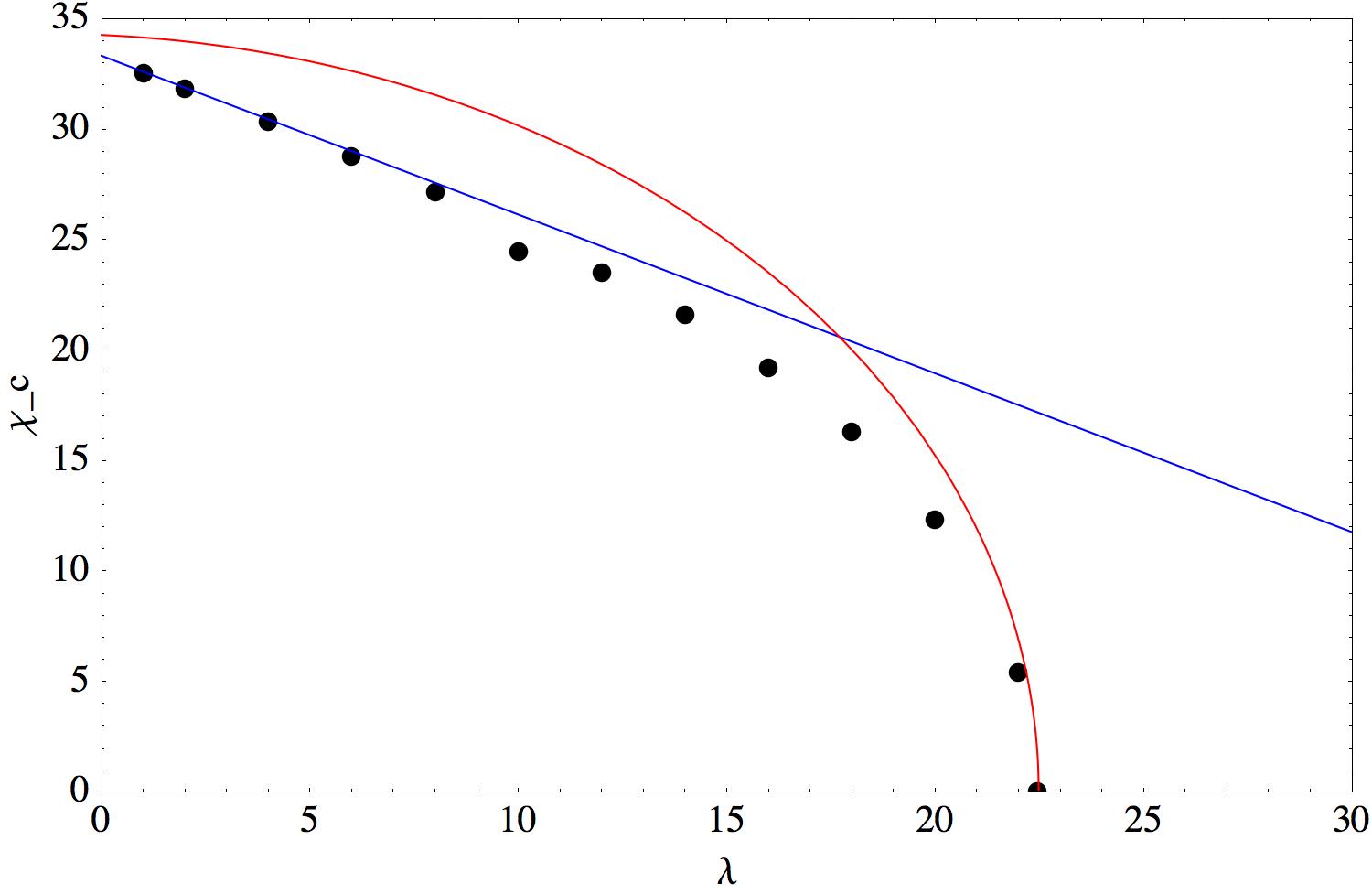}}
\begin{picture}(0,0)(0,0)
\put(70,70){$\chi_c$}
\put(170,0){$\visrat$}
\put(150,80){SW}
\put(200,120){\mbox{TB or intermittency}}
\end{picture}
      \caption{\footnotesize   Critical elasticity number vs viscosity ratio. The points denote transition from SW to intermittent or TB behavior as determined from the return maps. The red line is the upper boundary \refeq{up b}, $\hat\chi_c^{-1}=\sqrt{1-\hat\lambda^2}$, and the blue line is the \cite{Kessler-Finken-Seifert:2009}'s  result  $\hat\chi_c^{-1}= 1-\hat\lambda/2$ ; the  notation is $\hat\lambda=\sqrt{{\Delta}/{30\pi}}{(23\visrat+32)}/{8}$ and $\hat\chi^{-1}=2\chi\sqrt{{\Delta}/{30\pi}} $}
      \label{figChic}
\end{figure}
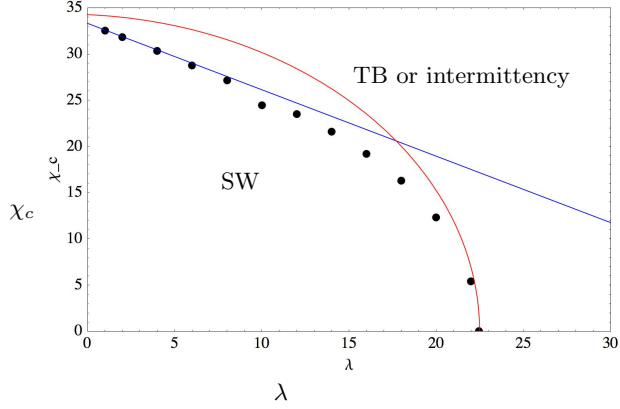

In order to find the phase boundary between the SW and TB, the angle evolution equation \refeq{psidot-el} is rewritten as
\begin{equation}
\label {psidot-transf}
\begin{split}
\dot \psi = -\frac{1}{2} + \frac{h}{\sqrt{\triangle}} A(\phi)\cos(2\psi+\delta(\phi)),
\end{split}
\end{equation}
where
\begin{equation}
\begin{split}
A = \left[1+\frac{4\triangle}{30\pi}\chi^2 + 4\chi\sqrt{\frac{\triangle}{30\pi}}\sin(2\phi)\right]^{1/2}\,,\quad \delta = \tan^{-1}\left(\frac{-2\sqrt{\frac{\triangle}{30\pi}}\chi\cos(2\phi)}{1+2\sqrt{\frac{\triangle}{30\pi}\chi\sin(2\phi)}}\right)\,.
\end{split}
\end{equation}
At the saddle-node bifurcation from SW to TB,
\[
\int^{2\pi}_0 \dot\psi dt = \psi(2\pi)-\psi(0) = 0\,.\]
Since $\cos(2\psi + \delta)$ is bounded between $-1$ and $+1$, 
 the integrand, given by \refeq{psidot-transf},  
would be zero if
\[\frac{h}{\sqrt{\triangle}} A^* =\frac{1}{2}.\]
In order to estimate the critical value of the elastic capillary number $\chi_c$, we observe that the  function $A(\chi)$ has a minimum when 
\begin{eqnarray}
\sin(2\phi) &=& \min\left\{-2\sqrt{\frac{\triangle}{30\pi}}\chi\,, -1\right\}\,.
\end{eqnarray}
At the bifurcation $A_{\mbox{min}}=A^*$,
which leads to (assuming that $\chi<\sqrt{30\pi/4\Delta})$
\begin{eqnarray}
\label{up b}
\chi_c &=& \sqrt{\frac{30\pi}{4\triangle}}\sqrt{1-\frac{\triangle}{4}\frac{1}{h^2}}.
\end{eqnarray}
Note that the above derivation also indicates that this boundary should be an upper boundary.
Figure \ref{figChic} shows the phase boundary between SW and TB/intermittent dynamics. 

\subsubsection{Weak elasticity}
In order to make further progress analytically, let us consider the limit of weak elasticity, $\chi\ll 1$.
In this case, the effect of shear elasticity is  analyzed as a small perturbation around the stationary fixed point corresponding to a tank-treading vesicle, $\psi(t)=\psi(\chi=0)+s(t)$, where $\psi(\chi=0)=1/2 \arccos\left(\sqrt{\Delta}/2h\right)$.
After linearization of \refeq{psidot-el} we obtain a linear forced oscillator equation
\begin{equation}
\label{exp osc}
\dot s +\left(\frac{2h}{\sqrt{\Delta}}\right) s=\left(\chi\frac{2h}{\sqrt{30\pi}}\right) \sin t\,.
\end{equation}
It can be integrated to give for the swinging motion
\begin{equation}
\label{small vis Elas}
s(t)=\alpha\cos(t+\beta)\,,
\end{equation}
where 
\begin{equation}
\beta=\frac{h}{\sqrt{\Delta}}\,,\quad \alpha=\frac{16 \chi}{\sqrt{(23 \visrat+32)^2+4\left(\frac{30\pi}{\Delta}\right)^2}}\,.
\end{equation}
Interestingly, the phase angle $\beta$ does not depend on the elasticity. The amplitude of the oscillations increases linearly with the elasticity number, i.e., increases with increasing elasticity or decreasing shear rate. Also, at a given elasticity number $\chi$,  the oscillations decrease with viscosity ratio.

Note that the above analysis is limited to  viscosity ratios such that the inclination angle $\psi(\chi=0)$ is not small, i.e. $\visrat\sim O(1)$. Only in this case, during the linearization we can neglect a term of the type $s\chi$  but retain the term $s\sin\psi(\chi=0) $ .  

\subsection{Effect of the deformability: analysis of the full system}


The deformability introduces  another parameter to the phase digram - the initial condition for the shape, $r_0$, which measures the asphericity along the flow-vorticity direction. 

In the previous section, we saw that the initial condition $g^0_{20}=0$  and  $r_0\ne 0$ imply constant  $r_0=\sqrt{\Delta}/2$ and the shape is a fixed ellipsoid whose major axis lies in the plane perpendicular to the undisturbed vorticity. In the limit of high resistance to shearing $\chi\rightarrow\infty$, the particle motion is given by the    $C = \infty$  Jeffrey orbit \cite[]{Jeffrey} .
 
If $g^0_{20}\ne 0$ but $r_0=0$ the capsule attains a stationary ellipsoid shape. In the rigid ellipsoid limit,  $\chi\rightarrow\infty$,  the motion is equivalent to the $C = 0$ Jeffery orbit (spinning around the axis of symmetry, which is parallel to the undisturbed vorticity ); the shape is axisymmetric and characterized by $f_{20}=\sqrt{\Delta/2}$ and $ f_{2\pm2}=0$.
For finite $\chi$ the shape is a general ellipsoid.

The most important consequence of the deformability is that it  suppresses the intermittency.  It appears that the intermittency is an artifact of  the reduced model. Once the full set of equations \refeq{psidot1} and \refeq{Rdot1} is solved, no intermittency is observed, only very long transients. This is in agreement with the numerical simulations by \cite{Kessler-Finken-Seifert:2008}, and in contrast to the conclusion by \cite{Noguchi:2009b} that  deformability does not change qualitatively the dynamics. The latter analysis, however,
ignores  the  effect of the tension due to the the area constraint.

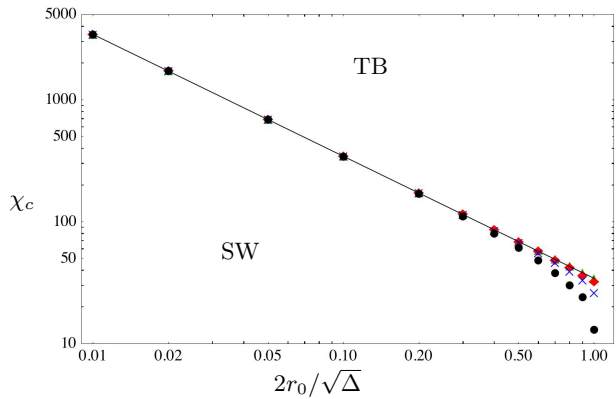
\begin{figure}
\centerline{\includegraphics[width=3in]{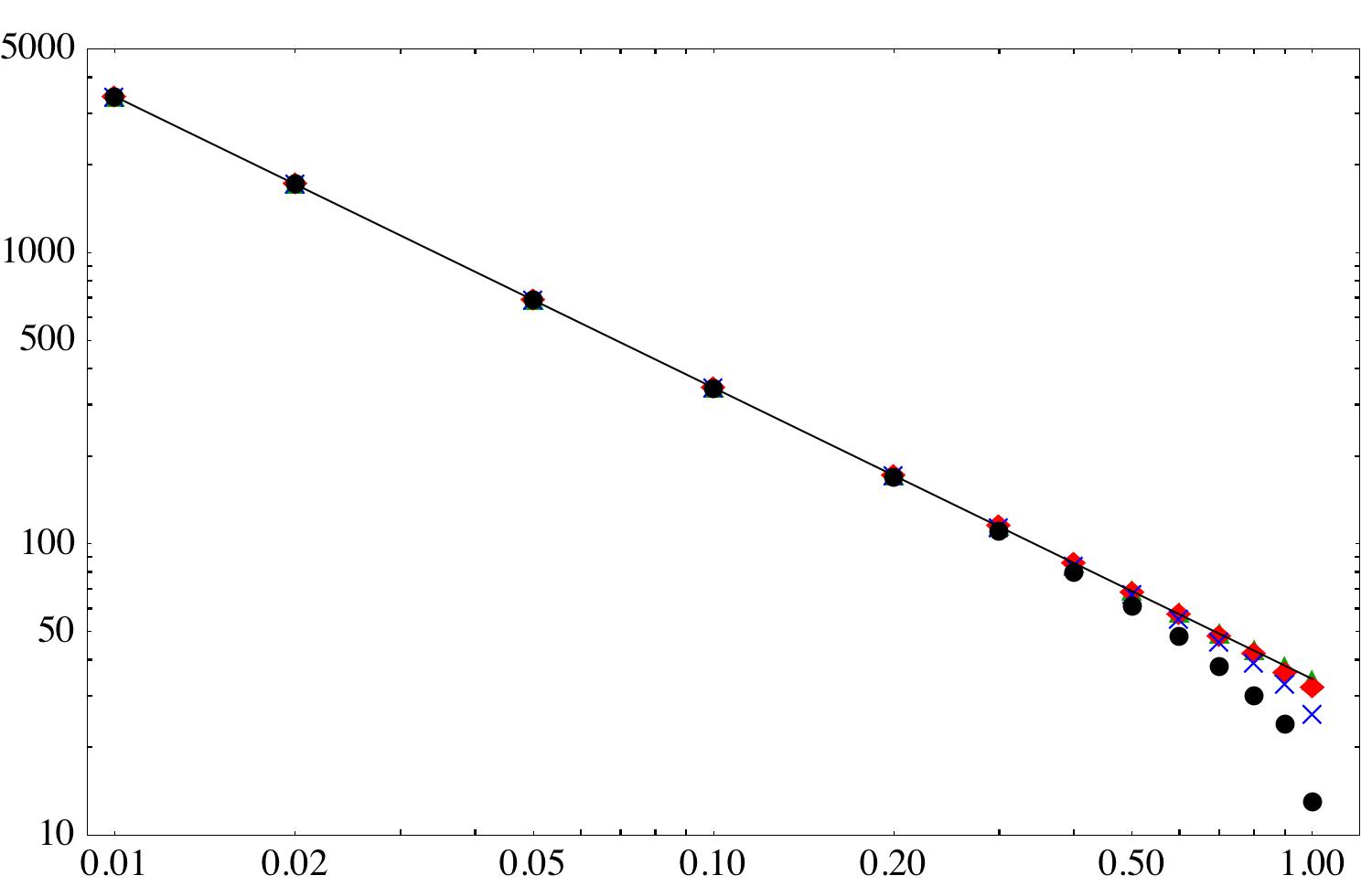}}
\begin{picture}(0,0)(0,0)
\put(70,70){$\chi_c$}
\put(170,0){$2r_0/\sqrt{\Delta}$}
\put(150,50){SW}
\put(200,120){TB}
\end{picture}
      \caption{\footnotesize Phase boundary between swinging and tumbling. Critical elasticity as a function of $2r_0/\sqrt{\Delta}$ for different viscosity ratios  $\visrat=0$ (green), $\visrat=5$ (red diamonds), $\visrat=10$ (blue crosses), $\visrat=20$ (black circles). $\Delta=0.02$. The line is $~1/r_0$}
      \label{fig6a}
\end{figure}
\begin{figure}
\centerline{\includegraphics[width=3in]{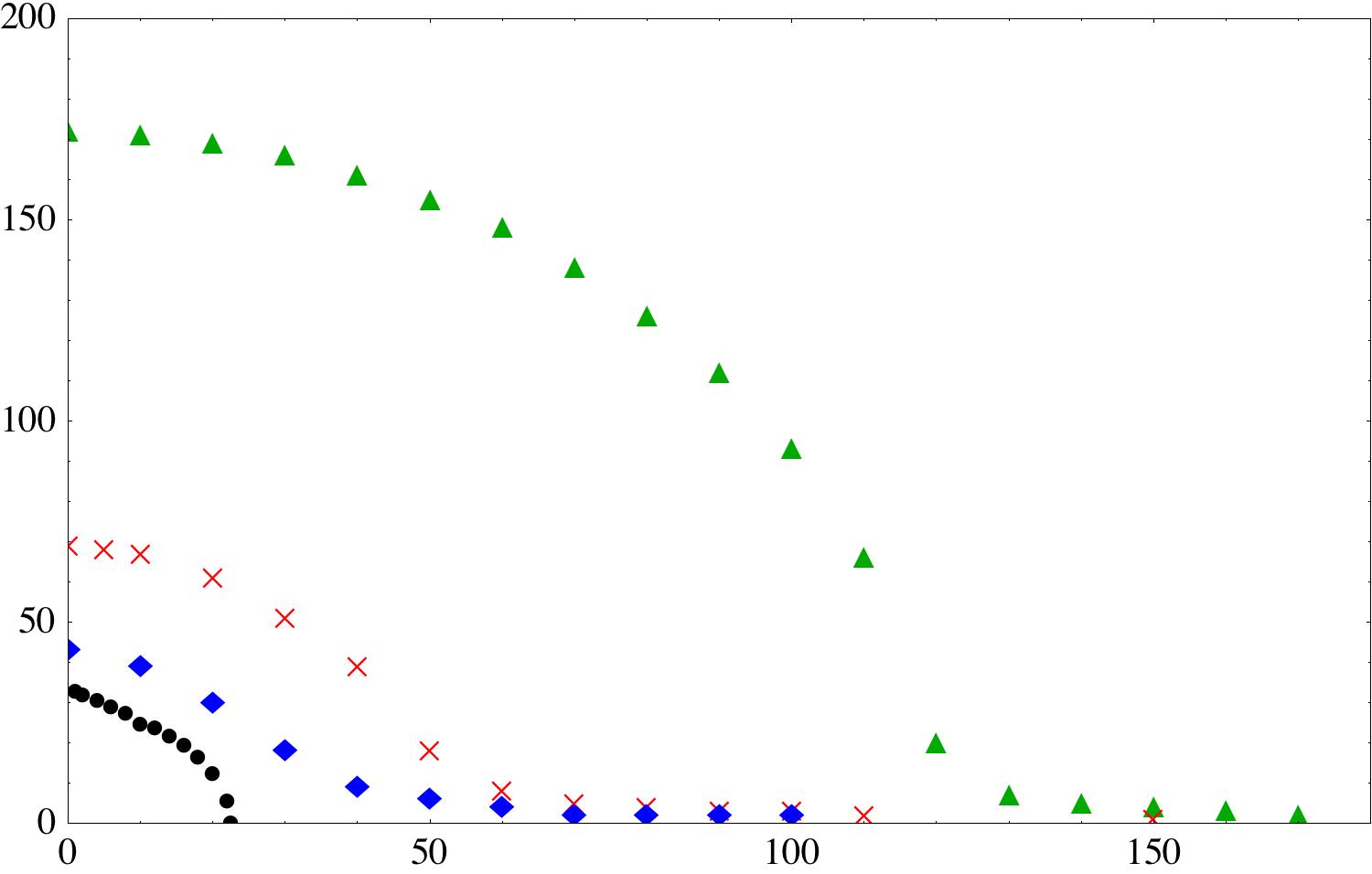}}
\begin{picture}(0,0)(0,0)
\put(70,70){$\chi_c$}
\put(170,0){$\visrat$}
\put(90,25){SW}
\put(200,120){TB}
\end{picture}
      \caption{\footnotesize Phase boundary between swinging and tumbling. Critical elasticity as a function of viscosity ratio for different deformability  $2r_0/\sqrt{\Delta}$:  0.2 (green triangles) , 0.5 (red crosses), 0.8 (blue diamonds), 1 (black circles). $\Delta=0.02$.}
      \label{fig6b}
\end{figure}
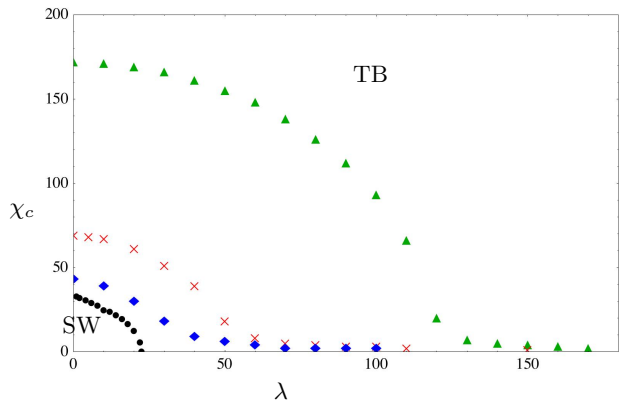
\begin{figure}
\centerline{\includegraphics[width=3in]{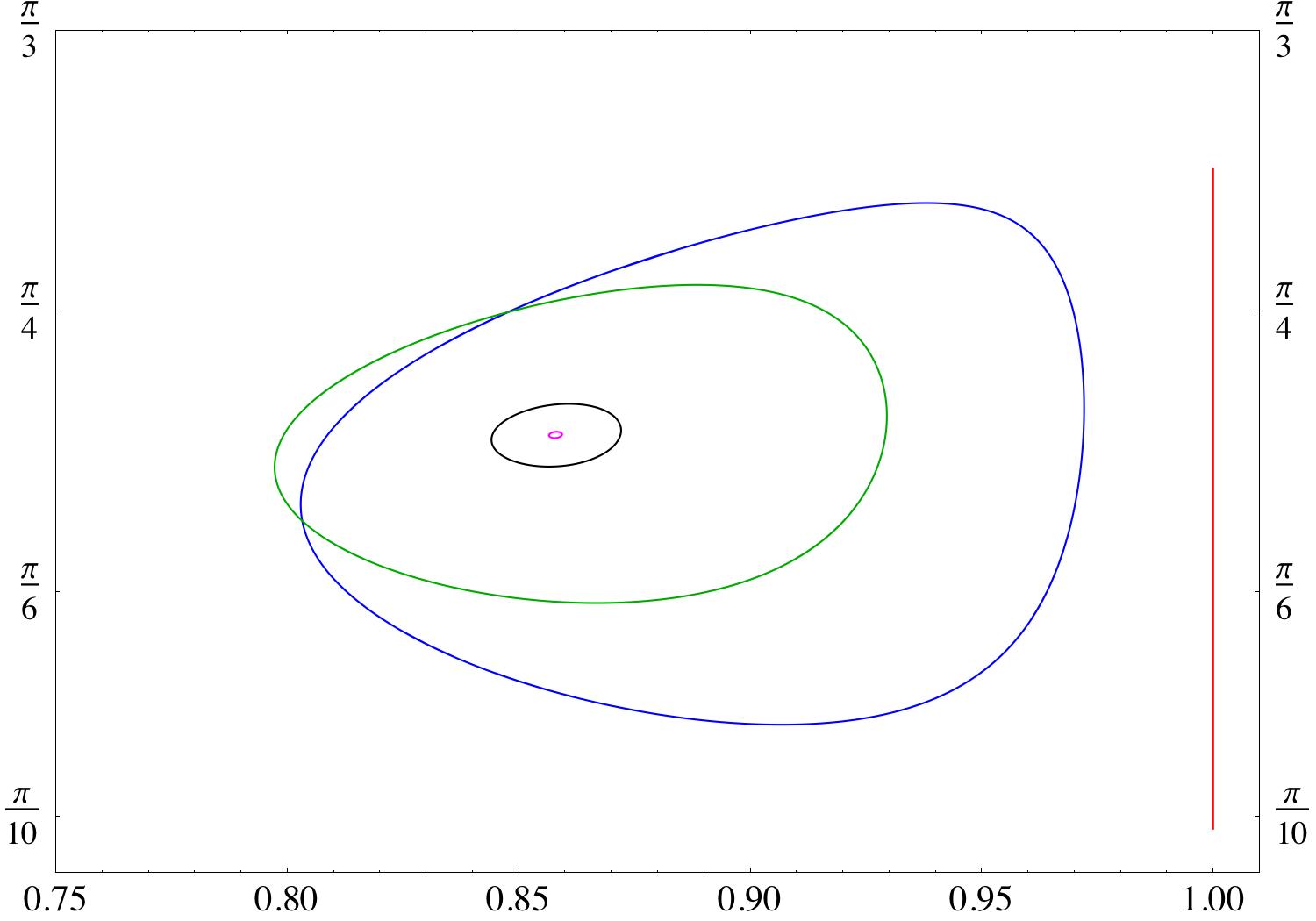}}
\begin{picture}(0,0)(0,0)
\put(70,70){$\psi$}
\put(170,0){$2R/\sqrt{\Delta}$}
\end{picture}
      \caption{\footnotesize Effect of $r_0$ on the limit cycle at fixed $\chi=20$, $\visrat=5$, $\Delta=0.02$. $2r_0/\sqrt{\Delta}=$ 1 (red), 0.8(blue), 0.5(green), 0.1(black) and 0.01(magenta) }
      \label{fig7a}
\end{figure}
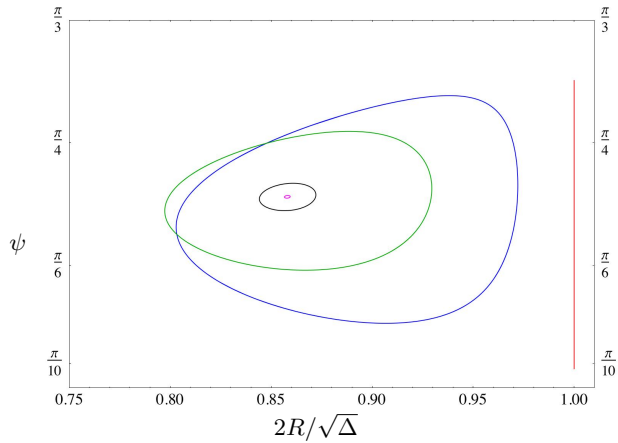
\begin{figure}
\centerline{\includegraphics[width=3in]{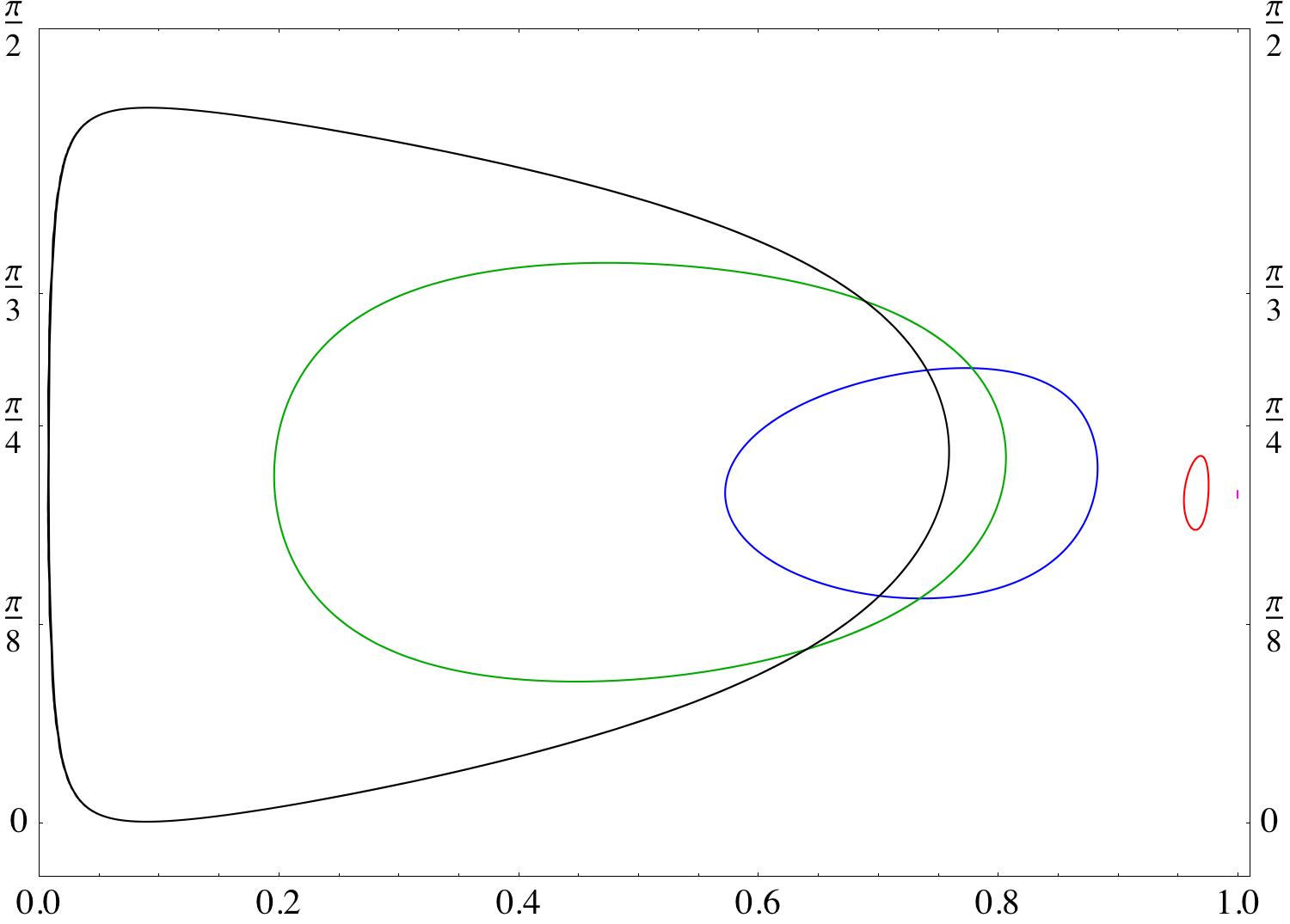}}
\begin{picture}(0,0)(0,0)
\put(70,70){$\psi$}
\put(170,0){$2R/\sqrt{\Delta}$}
\end{picture}
      \caption{\footnotesize Effect of $\chi$ on the limit cycle at given $2r_0/\sqrt{\Delta}=0.5$, $\visrat=5$, $\Delta=0.02$. $\chi=$  1(magenta), 10 (red), 30(blue), 50(green), 67(black).  }
      \label{fig7b}
\end{figure}
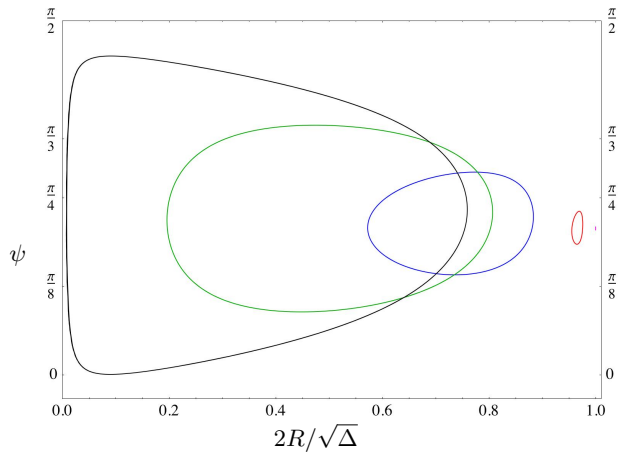
\begin{figure}
\centerline{\includegraphics[width=3in]{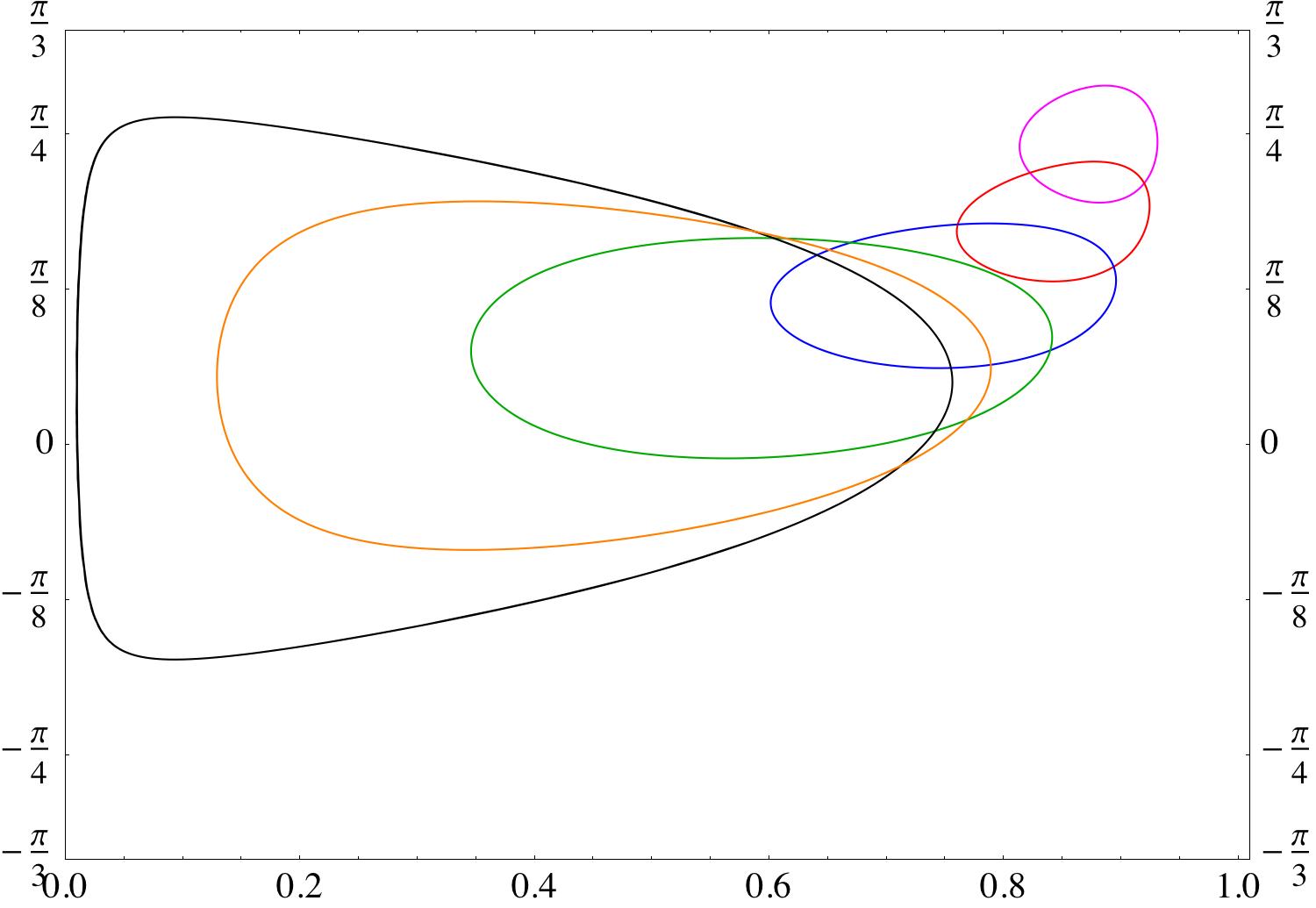}}
\begin{picture}(0,0)(0,0)
\put(70,70){$\psi$}
\put(170,0){$2R/\sqrt{\Delta}$}
\end{picture}
      \caption{\footnotesize Effect of $\visrat$ on the limit cycle at given $2r_0/\sqrt{\Delta}=0.5$, $\chi=20$, $\Delta=0.02$. $\visrat=$  0(magenta),10 (red), 20 (blue), 30(green), 40(orange), 48(black).  }
      \label{fig7c}
\end{figure}

The phase diagram for a deformable particle is presented in Figures \ref{fig6a} and  \ref{fig6b}. Figure \ref{fig6a} shows that as $r_0\rightarrow 0$, the critical elasticity number diverges as $1/r_0$.
Decreasing $r_0$, which allows for more deformation, widens the swinging region, as seen in Figure \ref{fig6b}.
Figure \ref{fig7a} illustrates the evolution of the limit cycle in the $R-\psi$ phase space.  The decreasing radius with increasing $r_0$ indicates that enhanced deformability decreases the amplitude of the oscillations.
Figure \ref{fig7b} shows that increasing the elasticity number (decreasing the shear rate) $\chi$ increases the amplitude of the oscillations, and eventually causes tumbling.
Finally, Figure \ref{fig7c} demonstrates  that increasing the viscosity ratio $\visrat$ increases the amplitude of the oscillations.

\section{Conclusions}

We have considered the dynamics of a deformable membrane--encapsulated fluid  particle in steady shear flow.  We have developed an analytical solution in the asymptotic case where the deformation is limited by small excess area. The theory accounts for the membrane area-incompressibility, resistance to shearing, and ellipsoidal unstressed shape.  The solution of the evolution equation  allowed us to construct the  phase diagram and to derive analytical results for the phase boundaries.

Our analysis clarifies the physical basis and limitations of earlier phenomenological models, which are restricted to fixed shape and compressible membrane.  In particular, we show that the reported intermittency is an artifact of the shape-preservation.

\section{Acknowledgement}

PMV acknowledges partial  financial support by NSF grant CBET--0846247. YNY thanks Roy Goodman for useful discussions and acknowledges financial support by NSF grants CBET--0853673 and DMS--0708977.

\appendix

\section{Relation to the analysis of \cite{Kessler-Finken-Seifert:2009}}
\label{kessler}

In Kessler et al.'s paper \cite{Kessler-Finken-Seifert:2009}, angles $\Sigma$ and $\Delta$ (not to be confused by the excess area in our notation) were employed. They are related to the inclination angle $\psi$ and the tank-trading angle $\phi$ as follows
\begin{equation}
\Sigma=-\phi-\frac{\pi}{4}\,,\quad \Delta=\phi-2\psi+\frac{\pi}{4}\,.
\end{equation}
Starting from their equations (21) and (22)
\begin{equation}
\label{keq}
\begin{split}
\partial_\tau \Sigma=&\hat\lambda\,\\
\partial_\tau\Delta=&\hat \lambda+4\sin\Sigma\sin\Delta+2\left(\hat \chi^{-1}-1\right)\cos\left(\Sigma-\Delta\right)
\end{split}
\end{equation}
we recover our equations \refeq{psidot1} and  \refeq{Rdot1},  if we identify $\lambda$ and $\chi^{-1}$   from their notation with
\begin{equation}
\begin{split}
\hat \lambda\Leftrightarrow \,\sqrt{\frac{\Delta}{30\pi}}\frac{23\visrat+32}{8}\,,\hat \chi^{-1}\Leftrightarrow \, 2\chi\sqrt{\frac{\Delta}{30\pi}}\,.
\end{split}
\end{equation}
Performing the same analysis on \refeq{keq} as in section \ref{phaseb} yields the phase boundary reported by Kessler et al., namely
\begin{eqnarray}
\chi_c &=& \frac{\sqrt{30\pi}}{2\sqrt{\triangle}}\left[1-\frac{\sqrt{\triangle}}{16\sqrt{30\pi}}(23\visrat+32)\right].
\end{eqnarray}

\section{Shape evolution equations for the deformation of an ellipsoidal particle in shear flow}
Ellipsoidal deformation is characterized by only  $j=2$ modes. Let us introduce $f_{jm}=f_{jm}'+\im f_{jm}^{''}$ and $f_{j-m}=(-1)^m(f_{jm}'-\im f_{jm}^{''})$; same for the $g_{jm}$


The shape evolution is described by
\begin{equation}
\label{ev1}
\frac{\partial f_{20}}{\partial t}=4 \frac{h}{\Delta} f_{20}f_{22}^{''}+\chi\frac{4  h}{\Delta\sqrt{30 \pi}}\left[g_{20}\left(\Delta-2 f_{20}^2\right)-4 f_{20}\left(g_{22}^{'}f_{22}^{'}+g_{22}^{''}f_{22}^{''}\right)\right]
\end{equation}
\begin{equation}
\frac{\partial f'_{22}}{\partial t}=-f_{22}^{''}+4 \frac{h}{\Delta} f_{22}^{'}f_{22}^{''}+\chi\frac{4  h}{\Delta\sqrt{30 \pi}}\left[g_{22}^{'}\left(\Delta-4 \left(f_{22}^{'}\right)^2\right)-2 f_{22}^{'}\left(g_{20}f_{20}+2g_{22}^{''}f_{22}^{''}\right)\right]
\end{equation}
\begin{equation}
\frac{\partial f^{''}_{22}}{\partial t}=-h+f_{22}^{'}+4 \frac{h}{\Delta}\left(f_{22}^{''}\right)^2+\chi\frac{4  h}{\Delta\sqrt{30 \pi}}\left[g_{22}^{''}\left(\Delta-4 \left(f_{22}^{''}\right)^2\right)-2 f_{22}^{''}\left(g_{20}f_{20}+2g_{22}^{'}f_{22}^{'}\right)\right]
\end{equation}
\begin{equation}
\label{ev4}
\frac{\partial g^{'}_{22}}{\partial t}=-\omega g^{''}_{22}\quad \frac{\partial g^{''}_{22}}{\partial t}=\omega g^{'}_{22}\quad \frac{\partial g_{20}}{\partial t}=0
\end{equation}

Next we describe the solution.

\section{Formalism}
\subsection{Velocity fields and hydrodynamics stresses}

Velocity fields are described using basis sets of fundamental solutions of the Stokes equations \cite[]{Schmitz-Felderhof:1982, Cichocki-Felderhof-Schmitz:1988,Vlahovska:2009a}, $\bu^\pm_{jmq}$, defined in Appendix~\ref{Ap:velocity basis}:
\begin{subequations}
\label{velocity fields}
\begin{equation}
\begin{split}
\bv^{\out}(\br)=&\sum_{jmq} c^{\infty}_{jmq}\left[\bu^{+}_{jmq}(\br)-\bu^{-}_{jmq}(\br)\right]+\sum_{jmq}c_{jmq}\bu^{-}_{jmq}(\br)\,,
\end{split}
\end{equation}
\begin{equation}
\bv^{\ins}(\br)=\sum_{jmq}c_{jmq}\bu^{+}_{jmq}(\br)\,.
\end{equation}
\end{subequations}
On a sphere $\bu^{+}_{jmq}(r=1)=\bu^{-}_{jmq}(r=1)$ and the velocity fields given by \refeq{velocity fields} are continuous.

The hydrodynamic tractions exerted on a surface with a  normal vector $\bn$ are $\bn\cdot\bT$. They are expanded in vector spherical harmonics \refeq{vector harmonics}
\begin{equation}
\label{traction definition}
\trac\equiv\bn\cdot\bT=t_{jmq}\bS_{jmq}
\end{equation}
 In the particular case of a sphere characterized with a normal vector $\hat\br$, the
amplitudes of the viscous tractions  and the velocity field are linearly related
\begin{subequations}
\begin{equation}
t^\out_{jmq}=\sum_{q'}^{2}c^\infty_{jmq'}\left(\Theta_{q'q}^{+}-\Theta_{q'q}^{-}\right)+\sum_{q'}^{2}c_{jmq'}\Theta_{q'q}^{-}\,
\end{equation}
\begin{equation}
\label{HD trac}
t^\ins_{jmq}=\sum_{q'}^{2}c_{jmq'}\Theta_{q'q}^{+}\,
\end{equation}
\end{subequations}
where $\Theta_{q'q}^{\pm}$ are obtained from the  velocity fields \refeq{vel basis -}--\refeq{vel basis +} \cite[]{Blawzdziewicz-Vlahovska-Loewenberg:2000},
\begin{equation}
\label{tracT0+}
\tracT^{+}_{qq'}\left(j\right)=
\left(
\begin{array}{ccc}
2j+1 & 0 & -3\left(\frac{j+1}{j}\right)^\half \\
0 & j-1 & 0 \\
-3\left(\frac{j+1}{j}\right)^\half & 0 & 2j+1+\frac{3}{j} \\
\end{array}
\right)
\end{equation}
\begin{equation}
\label{tracT0-}
\tracT^{-}_{qq'}\left(j\right)=
\left(
\begin{array}{ccc}
-2j-1 & 0 & 3\left(\frac{j}{j+1}\right)^\half \\
0 & -j-2 & 0 \\
3\left(\frac{j}{j+1}\right)^\half & 0 & -2j-1-\frac{3}{j+1} \\
\end{array}
\right)
\end{equation}

\subsection{Boundary conditions: Stress balance}
\label{leading sol}


The stress balance in terms of spherical harmonics  reads
\begin{equation}
\label{stress bal2}
\trac^\out_{jmq} -\visrat \trac_{jmq}^\ins=\trac^{\mem}_{jmq}\,.
\end{equation}
Tangential stresses correspond to the $q=0,1$ components, and the normal stresses - to  $q=2$.
The hydrodynamic tractions are given by  \refeq{HD trac}.
 The membrane tractions are \cite[]{Vlahovska:2007,Seifert:1999}
\begin{equation}
t^\mem_{jmq}=\Ca^{-1}_\kappa(t^\kappa_{jmq}+t^\sigma_{jmq})+\chi t^\el_{jmq}\,,
\end{equation}
where the bending  stresses are
\begin{equation}
t^\kappa_{jm2}=\textstyle j(j+1)\left(j-1\right)\left(j+2\right)f_{jm} \,, \qquad
t^\kappa_{jm0}=0\,,
\end{equation}
and the stresses due to membrane tension are
\begin{equation}
t^\sigma_{jm2}=\textstyle 2\sigma_{jm}+\sigma_0\left(j-1\right)\left(j+2\right)f_{jm}\,, \qquad
t^\sigma_{jm0}=-\sqrt{j(j+1)}\sigma_{jm}
\,.
\end{equation}
Note that the membrane tension is non-uniform under non-equilibrium conditions
\begin{equation}
\sigma=\sigma_0+\sum_{jm} \sigma_{jm}Y_{jm}\,.
\end{equation}
The surface elastic stresses have only in-plane shearing component
\begin{equation}
t^\el_{jm2}=0\,,\qquad
t^\el_{jm0}=2\left(j-1\right)\left(j+2\right)[j(j+1)]^{-1/2}(f_{jm}-g_{jm})
\end{equation}
where $g_{jm}$ denotes the reference configuration. Membrane stresses do not involve a $q=1$ component.

\section{Solution}
\label{app:solution}
\subsection{External flow}
\label{ext flow}

Simple shear flow is defined as
\begin{equation}
\label{inf_flow}
\bv^\infty= y{\bf{\hat x}}
\end{equation}
which translates into
\begin{equation}
\bv^\infty=\sum_{j=1}^2\sum_{m=-j}^{j}\sum_{q=0}^{2} c_{jmq}^\infty \bu^+_{jmq}
\end{equation}
\begin{equation}
\label{shearinf}
c^\infty_{2\pm20}=\mp \im \sqrt{\frac{\pi }{5}}\,,\,
c^\infty_{2\pm22}=\mp  \im \sqrt{\frac{2 \pi }{15}} \,,\,
c^\infty_{101}=-\im  \sqrt{\frac{2 \pi }{3}}
\end{equation}

\subsection{Rotation: tank-treading frequency}
First, we find the velocity field amplitude $c_{jm1}$ using the tangential stress balance with $q=1$
\begin{equation}
\trac^\out_{jm1} -\visrat\trac_{jm1}^\ins=0\,.
\end{equation}
This gives
\begin{equation}
c_{jm1} =c^{\infty}_{jm1}\frac{2j+1}{2+j+\lambda(j-1)}
\end{equation}
The imposed shear flow \refeq{shearinf} has a rotational component only with $j=1$ and hence the above relation reduces to
\begin{equation}
c_{101}=c^{\infty}_{101}
\end{equation}
i.e., the rigid body rotation is unperturbed by the particle, and the particle rotates with the same rate as the flow.

\subsection{Membrane incompressibility}
The
local area conservation implies that the velocity field at the
interface is solenoidal \cite[]{Seifert:1999}
\begin{equation}
\label{membrane incompressibility}
\bnabla_s \cdot \bv=0\,\quad {\mbox{at}}\quad r=1
\end{equation}
Therefore the amplitudes of the velocity field \refeq{velocity fields} are related
\begin{equation}
\label{solen}
c_{jm0}=\frac{2}{ \sqrt{j(j+1)}}c_{jm2}\,.
\end{equation}

\subsection{Tension}
The non-uniform part of the membrane tension, $\sigma_{jm}$, is determined from the tangential component  of the stress balance \refeq{stress bal2}, $q=0$,
\begin{equation}
\label{tension}
\begin{array}{ll}
\sigma_{jm}=\Ca^{-1}_\kappa\left[- c^{\infty}_{jm0}\frac{2\left(1+2j\right) }{\sqrt{j(j+1)}}+ c^\infty_{jm2}\frac{3(2j+1)}{j(j+1)} + c_{jm0}\frac{\left(2+j+\visrat\left(j-1\right)\right)}{2\sqrt{j(j+1)}}
+2\chi\frac{(j-1)(j+2)}{j(j+1)}(f_{jm}-g_{jm})\right]\,.
\end{array}
\end{equation}

\subsection{Normal velocity and shape evolution}
The tension \refeq{tension} is substituted into the normal component of the stress balance \refeq{stress bal2}, $ q=2$,  to obtain the normal velocity $c_{jm2}$
\begin{equation}
\label{ca}
c_{jm2}=C_{jm}-(j+2)(j-1){d(\visrat,j)}^{-1}\left[\Ca^{-1}(j(j+1)(j(j+1)+\sigma_0))f_{jm}+4 \chi (f_{jm}-g_{jm})\right]\,,
\end{equation}
where
\begin{equation}
\label{Cjm}
\begin{array}{ll}
C_{jm}=d(\visrat, j)^{-1}\left[c^\infty_{jm0}\sqrt{j(j+1)}\left(2j+1\right)+c^\infty_{jm2}\left(4j^3+6j^2-4j-3\right)\right]
\end{array}
\end{equation}
and
\begin{equation}
d(\visrat, j)=(4+3j^2+2j^3)+(-5+3j^2+2j^3)\visrat\,.
\end{equation}

Finally, the motion of the interface is determined from the kinematic condition
\begin{equation}
\label{interface evolution:2}
\frac{\partial f_{jm}}{\partial t}= c_{jm2}+\bv_s \cdot \nabla f\quad {\mbox{at}}\,\, r=1\,.
\end{equation}
where $\bv_s=\bv^{\out}(r=1)=\bv^{\ins}(r=1)$.
At leading order, it takes the form
\begin{equation}
\label{interface evolution:3}
\frac{\partial f_{jm}}{\partial t}= c_{jm2}+\textstyle{\im \omega \frac{m}{2} }f_{jm}\quad {\mbox{at}}\,\, r=1\,.
\end{equation}
$\omega=1$ is the local rate of rotation. Note that $\frac{\partial }{\partial t}-\im \omega {m}/{2}$  is in fact the Jaumann derivative.
Substituting $c_{jm2}$ in \refeq{interface evolution:2} yields the evolution equation  for the  shape parameters
\begin{equation}
\label{ev equation f}
\frac{\partial f_{jm}}{\partial t}=\textstyle{\im \omega\frac{ m}{2}}f_{jm}+C_{jm}+\Ca^{-1}_\kappa\left(\Gamma_1+\sigma_0 \Gamma_2\right)f_{jm}+\chi \Gamma_3 (f_{jm}-g_{jm})+O\left(\Delta\right)\,,
\end{equation}
where 
\begin{equation}
\label{Gamma1}
\Gamma_1=-(j+2)(j-1)\left([j(j+1)]^2\right){d(\visrat,j)}^{-1}\,,
\end{equation}
\begin{equation}
\label{Gamma2}
\Gamma_2=-(j+2)(j-1)j(j+1){d(\visrat,j)}^{-1}\,,
\end{equation}
\begin{equation}
\label{Gamma3}
\Gamma_3=-4(j+2)(j-1){d(\visrat,j)}^{-1}\,,
\end{equation}
The reference configuration simply rotates with the flow
\begin{equation}
\label{ev equation g}
\frac{\partial g_{jm}}{\partial t}=\textstyle{\im \omega\frac{m}{2}}g_{jm}\,.
\end{equation}

\subsection{Isotropic tension}
The normal velocity \refeq{ca} and the shape evolution \refeq{ev equation f} include the yet unknown  isotropic membrane tension.
It is expressed  in terms of the shape modes and other known parameters in the problem using the area constraint \cite[]{Vlahovska:2007}
\begin{equation}
\label{delta to second order}
\Delta=\sum_{jm}a(j)f_{jm}f^*_{jm}\,, \quad a(j)=\frac{\left(j+2\right)\left(j-1\right)}{2}\,.
\end{equation}
Since $\dot\Delta=0$, it follows that
\begin{equation}
\sum a(j)\left(f^*_{jm}\frac{\partial f_{jm}}{\partial t}+f_{jm}\frac{\partial f^*_{jm}}{\partial t}\right)=0\,.
 \end{equation}
Using \refeq{ev equation f} leads to
\begin{equation}
\label{tension0}
\begin{split}
\sigma_0=-\frac{\Ca_\kappa} {\sum_{jm} a(j) \Gamma_2 f_{jm}f^*_{jm}}\sum_{jm} a(j) \left[C_{jm}f^*_{jm}+ \Ca^{-1}_\kappa \Gamma_1 f_{jm}f^*_{jm}+\chi \Gamma_3 (f_{jm}-g_{jm})f^*_{jm}\right]\,.
\end{split}
\end{equation}

In order to clarify the physical significance  of the isotropic tension, let us consider  the particular case when only  the ellipsoidal deformation modes, $j=2$, are present. \refeq{tension} simplifies to
\begin{equation}
\sigma_0=-6-\im \frac{\Ca_\kappa}{\Delta}\sqrt{\frac{10 \pi}{3}}\left(f_{22}-f_{2-2}\right)+\frac{2}{3}\chi \Ca_\kappa \left[-1+\frac{2}{\Delta}\left(g_{20} f_{20}+g_{22} f_{2-2}+g_{2-2} f_{22} \right)\right]
\end{equation}
We see that the tension varies with deformation. 

Inserting the tension \refeq{tension0} into the shape evolution \refeq{ev equation f} and keeping only the $j=2$ modes yields \refeq{ev1}-\refeq{ev4}.

\section{Spherical harmonics}
\label{Harmonics}

A good reference on spherical harmonics is \cite{Varshalovich:1988}.
The normalized spherical scalar harmonics are defined as
\begin{equation}
\label{normalized spherical harmonics}
   Y_{jm}\left(\theta,\varphi\right) = \textstyle \left[\frac{2j+1}{4\pi}\frac{(j-m)!}{(j+m)!}\right]^\half (-1)^m P_j^m(\cos\theta)e^{{\rm i}m\varphi},
\end{equation}
where  $\rhat=\br/r$,  $(r, \theta,\varphi)$
are the spherical coordinates, and $P_j^m(\cos\theta)$ are the Legendre polynomials.
For example
\begin{equation}
Y_{10}=\frac{1}{\sqrt{4 \pi}}\cos\theta \,.
\end{equation}
The vector spherical harmonics 
are defined as \cite{Blawzdziewicz-Vlahovska-Loewenberg:2000}
\begin{equation}
\label{vector harmonics}
\begin{split}
\bS_{jm0}&=\left[j\left(j+1\right)\right]^{-\half}r\nabla_\Omega Y_{jm}\,,\\
 \bS_{jm2}&=\rhat Y_{jm}\,,\\
\quad \bS_{jm1}&=-\im \rhat \times \bS_{jm0}
\end{split}
\end{equation}
where $\nabla_{\Omega}$ denotes the angular part of the gradient operator.

\section{Fundamental set of velocity fields}
\label{Ap:velocity basis}
We reproduce the velocity and stress fields from \cite{Vlahovska:2007}. First we list the expressions for the functions ${\bu}^\pm_{jmq}\left(r, \theta,\varphi\right)$. The velocity field outside the vesicle is described by
\begin{subequations}
\label{vel basis -}
\begin{equation}
\label{-vel 0}
\begin{array}{ll}
\bu^-_{jm0}=&{\textstyle\frac{1}{2}}r^{-j}\left(2-j+j r^{-2}\right)\bS_{jm0}+{\textstyle\frac{1}{2}}r^{-j}\left[j\left(j+1\right)\right]^{1/2}\left(1-r^{-2}\right) \bS_{jm2}\,,
\end{array}
\end{equation}
\begin{equation}
\label{-vel 1}
\begin{array}{ll}
\bu^-_{jm1}=\textstyle r^{(-j-1)} \bS_{jm1}\,,
\end{array}
\end{equation}
\begin{equation}
\label{-vel 2}
\begin{array}{ll}
\bu^-_{jm2}=&{\textstyle\frac{1}{2}}r^{-j}\left(2-j\right) \left(\frac{j}{1+j}\right)^{1/2}\left(1-r^{-2}\right)\bS_{jm0}+{\textstyle\frac{1}{2}}r^{-j}\left(j+(2-j)r^{-2}\right)\bS_{jm2}\,.
\end{array}
\end{equation}
\end{subequations}
The velocity field inside the vesicle is described by
\begin{subequations}
\label{vel basis +}
\begin{equation}
\label{+vel 0}
\begin{array}{ll}
\bu^+_{jm0}=& {\textstyle\frac{1}{2}}r^{j-1}\left(-(j+1)+(j +3)r^2\right)\bS_{jm0}-{\textstyle\frac{1}{2}}r^{j-1}\left[j\left(j+1\right)\right]^{1/2}\left(1-r^2\right)\bS_{jm2}\,,
\end{array}
\end{equation}
\begin{equation}
\label{+vel 1}
\begin{array}{ll}
\bu^+_{jm1}=\textstyle r^j \bS_{jm1}\,,
\end{array}
\end{equation}
\begin{equation}
\label{+vel 2}
\begin{array}{ll}
\bu^+_{jm2}=&{\textstyle\frac{1}{2}}r^{j-1}\left(3+j\right)\left(\frac{j+1}{j}\right)^{1/2}\left(1-r^2\right)\bS_{jm0}+{\textstyle\frac{1}{2}}r^{j-1}\left(j +3-(j+1)r^2  \right)\bS_{jm2}\,.
\end{array}
\end{equation}
\end{subequations}
On a sphere $r=1$ these velocity fields reduce to the vector spherical harmonics defined by \refeq{vector harmonics}
\begin{equation}
\bu^{\pm}_{jmq}=\bS_{jmq}\,.
\end{equation}
Hence, $\bu^{\pm}_{jm0}$ and $\bu^{\pm}_{jm1}$ are tangential,  and $\bu^{\pm}_{jm2}$ is normal to a sphere. In addition,  $\bu^{\pm}_{jm0}$ defines an irrotational velocity field.

\bibliographystyle{jfm}
\bibliography{refs}
\end{document}